\documentclass[preprint,prb,12pt]{revtex4-1}
\usepackage{bm}
\usepackage[sumlimits]{amsmath}
\usepackage{xcolor}
\usepackage{makeidx}
\usepackage{graphicx}
\usepackage[hidelinks,colorlinks=true,linkcolor=black]{hyperref}
\usepackage{float}
\usepackage{bm}
\usepackage{amsmath,amssymb}
\usepackage[labelsep=period]{caption}
\usepackage{xspace}
\usepackage[nameinlink]{cleveref}
\usepackage{subcaption}

\newcommand*{\average}[1]{{\langle #1\rangle}}

\newcommand*{\floor}[1]{{\lfloor #1\rfloor}}
\newcommand*{\abs}[1]{|#1|}

\newcommand*{\set}[1]{\{ #1 \}}
\newcommand*{\Set}[1]{\left\{#1\right\}}
\newcommand*{\median}{\mathrm{median}}
\newcommand*{\Det}[1]{\left|#1\right|}
\newcommand*{\transpose}[1]{{#1}^{\mathrm{T}}}
\newcommand*{\inverse}[1]{{#1}^{-1}}
\newcommand*{\Kgauss}{K_{\mathrm{G}}}
\newcommand*{\Kconst}{K_{\mathrm{C}}}
\newcommand*{\kb}{k_{\mathrm{B}}}

\newcommand*{\Ndata}{N_{\mathrm{data}}}
\newcommand*{\Naverage}{N_{\mathrm{ave}}}
\newcommand*{\Nsize}{N_{\mathrm{size}}}

\newcommand*{\mprime}{\partial m}%{m^{\prime}}
\newcommand*{\meq}{m_{\mathrm{eq}}}
\newcommand*{\Tc}{T_{\mathrm{C}}}

\newcommand*{\ypredict}{Y_{\text{predict}}}
\newcommand*{\thetaOpt}{\bm{\theta}_{\text{opt}}}
\newcommand*{\vfinal}{v_{k, \mathrm{final}}}
\newcommand*{\tmin}{t_{\mathrm{min}}}
\newcommand*{\tmax}{t_{\mathrm{max}}}

\begin{document}

\author{Yuma Osada}
\email{osada.yum@gmail.com}
\affiliation{Department of Engineering Science, Graduate School of Informatics and Engineering, The University of Electro-Communications, 1-5-1 Chofugaoka, Chofu-shi, Tokyo 182-8585, Japan}
\author{Yukiyasu Ozeki}
\affiliation{Department of Engineering Science, Graduate School of Informatics and Engineering, The University of Electro-Communications, 1-5-1 Chofugaoka, Chofu-shi, Tokyo 182-8585, Japan}
\title{Improvement of Analysis for Relaxation of Fluctuations by Using Gaussian Process Regression and Extrapolation Method}
\date{\today}
\begin{abstract}
The nonequilibrium relaxation (NER) method, which has been used to investigate equilibrium phase transitions via their nonequilibrium behavior, has been widely applied to various models to estimate critical temperatures and critical exponents.
Although the estimation of critical temperatures has become more reliable and reproducible, that of critical exponents raises concerns about the method's reliability.
Therefore, we propose a more reliable and reproducible approach using Gaussian process regression.
In addition, the present approach introduces statistical errors through the bootstrap method by combining them using the extrapolation method.
Our estimation for the two-dimensional Ising model yielded \(\beta = 0.12504(6)\), \(\gamma = 1.7505(10)\), and \(\nu = 1.0003(6)\), which are consistent with the exact values.
The value \(z = 2.1669(9)\) is reliable because of the high accuracy of these exponents.
We also obtained the critical exponents for the three-dimensional Ising model and found that they are close to those reported in a previous study.
Thus, for systems undergoing second-order transitions, our approach improves the accuracy, reliability, and reproducibility of the NER analysis.
Because the proposed approach requires the relaxation of some observables from Monte Carlo simulations, its simplicity imparts it with significant potential.
\end{abstract}
\maketitle

\crefname{section}{Sect.}{Sects.}
\Crefname{section}{Section}{Sections}
\crefname{equation}{Eq.}{Eqs.}
\Crefname{equation}{Equation}{Equations}
\crefname{figure}{Fig.}{Figs.}
\Crefname{figure}{Figure}{Figures}
\crefname{table}{Table}{Tables}
\Crefname{table}{Table}{Tables}

\section{\NoCaseChange{Introduction}}
\label{sec:org0d2a36f}
Critical exponents are essential for understanding critical universality.
Various methods for estimating critical exponents have been developed, including nonequilibrium relaxation (NER) analysis,~\cite{ozeki1998multicritical,ozeki2003nonequilibrium,ozeki2007nonequilibrium} finite-size scaling analysis, \cite{harada2011bayesian,harada2015kernel} conformal-bootstrap theory,~\cite{kos2016precision} and the tensor-renormalization group method.~\cite{levin2007tensor}
The NER method is used to analyze the properties of a system's equilibrium state on the basis of its nonequilibrium behavior.
It has been applied to various systems.
In particular, it has been used to analyze slow-relaxing systems, such as those undergoing critical slowing down.
Examples include systems that undergo the Kosterlitz--Thouless~\cite{berezinskii1971destruction,berezinskii1972destruction,J_M_Kosterlitz_1973,J_M_Kosterlitz_1974} transition,~\cite{ozeki2007nonequilibrium,ozeki2019dynamical}, spin-glass systems,~\cite{ozeki2014numerical,ozeki1998multicritical,terasawa2023dynamical} and fully frustrated systems.~\cite{ozeki2020dynamical,osada2023dynamical}
In addition, the scheme of the NER method has been applied to a transition in the percolation model.~\cite{hagiwara2022size}
Thus, the NER analysis is applicable to numerous models and contributes to research on phase transitions and critical phenomena.

The difficulty in the NER analysis of fluctuations lies in calculating the differentiation of discrete values obtained from a simulation.
In the NER analysis of fluctuations, critical exponents are derived from the slope of the data values.
For the NER method, in contrast to the well-established dynamical scaling analysis for critical temperatures,~\cite{echinaka2016dca} a stable and automatic analysis to determine critical exponents has not yet been developed.
There are two primary issues with the conventional method used in this analysis.
The first issue is the unstable nature of the slopes produced by the conventional method.
Simple numerical differentiation is ineffective because of the noise present in the data.
For the conventional method, efforts have been made to reduce the effect of noise using a linear approximation to determine the slope.
However, the instability still needs to be addressed and the reliability of the analysis must be determined.
The second issue is the difficulty in determining the convergence behavior from discontinuous and non-monotonic values when extrapolating to estimate critical exponents.
In the conventional method, the form of the slope is assumed to be \(a_{1}(1/t)^{a_{2}} + a_{3}\), which emphasizes the short-time behavior.
Note that the time \(t\) is measured in a unit of Monte Carlo steps (MCSs) in simulations.
For example, the interval \(t = [1, 10]\) corresponds to \(1/t = [0.1, 1]\), which is \(90\%\) of the interval \(1/t = [0, 1]\)(\(t = [1, \infty]\)).
These drawbacks undermine the reliability of extrapolations to the thermodynamic limit as \(t \to \infty\).
To overcome these issues, we use Gaussian process regression for the continuous slope.
Selecting an appropriate extrapolation method based on this continuous slope is critical.

Through this work, we aim to enhance the reliability and reproducibility of analyses by offering a systematic approach to the NER analysis of fluctuations.
We demonstrate the utility of the present method using two systems.
In the two-dimensional square Ising model, we validate our approach by comparing the critical exponents obtained at the exact critical temperature.
In the three-dimensional cubic Ising model, where the exact critical temperature remains unknown, we demonstrate the applicability of the present method near the critical temperature through comparisons with previous studies.

The remainder of this paper is organized as follows:
In \cref{sec:NER-explanation}, we explain the NER method of fluctuation.
In \cref{sec:GPregression-explanation}, we propose an improved analysis and validate our approach at the exact critical temperature using the two-dimensional Ising model.
In \cref{sec:application-3D-ising}, we apply the present method to the three-dimensional Ising model and justify it by comparing the results with those obtained using the methods reported in previous studies.
In \cref{sec:summary-discussion}, we summarize the present study and the proposed method.

\section{\NoCaseChange{Estimation of Critical Exponents Using Nonequilibrium Relaxation of Magnetization and Fluctuations}}
\label{sec:NER-explanation}
Let us explain how to use NER data to estimate the critical exponents \(\beta\), \(\gamma\), \(\nu\), and \(z\) of systems that undergo a second-order transition.
In the NER analysis, we simulate a large system considered to have no finite-size effects in the observed time interval.
We use
\begin{equation}
  \label{equ:magne-definition}
  m(t) \equiv \frac{1}{\Nsize}\sum_{i}s_{i}(t)
\end{equation}
as the dynamical order parameter at time \(t\), where \(s_{i}\) is \(\pm 1\) and \(\Nsize\) denotes the number of spins.
Hereafter, we measure the time \(t\) in simulations in a unit of MCSs.
The simulation aims to calculate relaxations of some quantities numerically, including the order parameter \(m(t)\) and its fluctuations, which asymptotically exhibit algebraic behavior with respect to time \(t\) at the critical temperature;~\cite{ozeki2007nonequilibrium}
\textit{e.g.}, the relaxation of magnetization \(m(t)\) from the all-up state shows an algebraic behavior, \(m~{(t)}~\sim~{t^{-\beta/(z\nu)}}\).
To observe the asymptotic power clearly from the estimated numerical data up to a finite maximum time, the local exponent of magnetization \(\lambda_{m}\) defined by
\begin{equation}
  \label{equ:lambda_m}
  \lambda_{m}(t) \equiv - \frac{\partial \log m(t)}{\partial \log t} \to \frac{\beta}{z\nu}
\end{equation}
is useful.
This local exponent is also useful in estimating critical temperatures because the asymptotic behavior of magnetization is represented by
\begin{equation}
  \label{equ:magnetization-temperature}
  m(t) \sim
    \begin{cases}
      \exp(- t / \tau)  \quad & T > \Tc,\\
      t^{- \beta / (z\nu)} \quad & T = \Tc,\\
      \meq              \quad & T < \Tc,\\
    \end{cases}
\end{equation}
where \(\tau\) is the relaxation time.
Although we can determine the range of critical temperatures from the relaxation of \(\lambda_{m}(t)\), which is called pinching estimation, we do not estimate critical temperatures in this study because we use the exact critical temperature for the two-dimensional Ising model and the highly accurate critical temperature obtained from massive calculations in the previous NER analysis reported by Ito~\cite{ito2005nonequilibrium} for the three-dimensional Ising model.
The relaxation of fluctuations, which asymptotically exhibits algebraic behavior with respect to time \(t\) at the critical temperature, is defined as
\begin{equation}
  \label{equ:fluctuation}
  \begin{aligned}
    \chi(t)
    &\equiv \average{m(t)^{2}} - \average{m(t)}^{2} \sim t^{\gamma / (z\nu)}\\
    \mprime(t)
    &\equiv \abs{\average{m(t)e(t)} - \average{m(t)}\average{e(t)}} \sim t^{(1 - \beta) / (z\nu)},\\
  \end{aligned}
\end{equation}
where \(\average{\cdot}\) denotes the dynamical average and \(e(t)\) represents the internal energy per site.
Note that we use variance fluctuations in the present paper, instead of the dimensionless fluctuations that have been used in previous NER studies,~\cite{ozeki2007nonequilibrium} because they enable easier calculation of errors of fluctuations.
The present method can be used for either variance fluctuations or dimensionless fluctuations, and the accuracy does not appear to change for either.
Their local exponents are also defined as
\begin{equation}
  \label{equ:lambda_fluctuations}
  \begin{aligned}
    \lambda_{\chi}(t)
    &\equiv \frac{\partial \log \chi(t)}{\partial \log t} \to \frac{\gamma}{z\nu}\\
    \lambda_{\mprime}(t)
    &\equiv \frac{\partial \log \mprime(t)}{\partial \log t} \to \frac{1 - \beta}{z\nu}.\\
  \end{aligned}
\end{equation}
From combinations of these local exponents, we can derive functions that asymptotically approach critical exponents:
\begin{equation}
  \label{equ:critical_exponents}
  \begin{aligned}
    \beta
    &= \lim_{t \to \infty} \beta(t) = \lim_{t \to \infty} \frac{\lambda_{m}(t)}{\lambda_{\mprime}(t) + \lambda_{m}(t)}\\
    \gamma
    &= \lim_{t \to \infty} \gamma(t) = \lim_{t \to \infty} \frac{\lambda_{\chi}(t)}{\lambda_{\mprime}(t) + \lambda_{m}(t)}\\
    \nu
    &= \lim_{t \to \infty} \nu(t) = \lim_{t \to \infty} \frac{2\lambda_{m}(t) + \lambda_{\chi}(t)}{d(\lambda_{\mprime}(t) + \lambda_{m}(t))}\\
    z
    &= \lim_{t \to \infty} z(t) = \lim_{t \to \infty}\frac{1}{\nu\left(\lambda_{\mprime}(t) + \lambda_{m}(t)\right)},\\
  \end{aligned}
\end{equation}
where \(d\) denotes the dimension of the system.
Consequently, we can estimate critical exponents by simulating \(m(t), \chi(t)\), and \(\mprime(t)\) at the critical temperature, differentiating them on a double-logarithmic scale, and extrapolating their combinations.

\section{\NoCaseChange{Improvement through Gaussian Process Regression}}
\label{sec:GPregression-explanation}
\subsection{Using Gaussian process regression}
\label{sec:org64c489d}
Let us first explain how to obtain local exponents in \cref{equ:lambda_m,equ:lambda_fluctuations} by differentiating values on a double-logarithmic scale.
In contrast to numerical differentiation, Gaussian process regression enables us to obtain analytic and continuous derivatives.~\cite{nakamura2016measurements,nakamura2020machine,johnson2020kernel}
We here briefly explain this process.
We aim to obtain the regression function for \(\Ndata\) data points \((X_{i}, Y_{i}, E_{i})\), where \(i\) is the label for data, \(X_{i}\) is an explanatory variable for \(Y_{i}\), and \(E_{i}\) represents the error in \(Y_{i}\) for \(i = 1, \ldots, \Ndata\).
We maximize the log-likelihood function by optimizing hyperparameters \(\bm{\theta}\).
The log-likelihood function for a Gaussian process is defined as
\begin{equation}
  \label{equ:log-likelihood}
  \log L(\bm{\theta}) = - \frac{1}{2}\log \Det{\Sigma(\bm{\theta})} - \frac{1}{2}\transpose{\bm{Y}} \inverse{\Sigma(\bm{\theta})} \bm{Y} - \frac{N}{2}\log(2\pi),
\end{equation}
where \(\Sigma(\bm{\theta})\) is the \(\Ndata \times \Ndata\) variance--covariance matrix and \(\Det{\Sigma}\) denotes the determinant of \(\Sigma\).
Element \((i, j)\) of \(\Sigma\) is defined by
\begin{equation}
  \label{equ:variance-covariance-matrix}
  \Sigma_{ij}(\bm{\theta}) = E_{i}^{2}\delta_{ij} + K(X_{i}, X_{j}, \bm{\theta}),
\end{equation}
where \(K(X_{i}, X_{j}, \bm{\theta})\) is a kernel function.
In Gaussian process regression, assuming all data points obey a multivariate Gaussian distribution, we can predict new points assumed to follow that distribution.
Specifically, we can predict \(Y\) at \(X\) using optimized hyperparameters \(\thetaOpt\) as
\begin{equation}
  \label{equ:prediction}
  \ypredict(X) = \transpose{\bm{k}} \inverse{\Sigma(\thetaOpt)} \bm{Y},
\end{equation}
where \(\bm{k} = (k_{i})\), \(k_{i}(X) = K(X_{i}, X, \thetaOpt)\), and \(\ypredict\) is the prediction function for \(Y\).
We can analytically obtain the derivative of \(Y\) by
\begin{equation}
  \label{equ:prediction-of-derivative}
  \frac{\partial \ypredict(X)}{\partial X} = \left(\frac{\partial \transpose{\bm{k}}}{\partial X}\right) \inverse{\Sigma(\thetaOpt)} \bm{Y}.
\end{equation}

In the following discussion, we demonstrate how to estimate critical exponents using the two-dimensional square Ising model.
The dynamical order parameter for this model, the magnetization \(m(t)\), is calculated as \(m(t) = {\Nsize}^{-1}\sum_{i}s_{i}(t)\), starting from the all-up state.
We analyze data pairs \(t_{i}\) versus \(y_{i}\) (\(= 1 / m_{i}, \chi_{i}\), or \(\mprime_{i}\)) to apply Gaussian process regression for differentiation.
In this paper, we use a composition kernel function consisting of a Gaussian kernel \(\Kgauss\) and a constant kernel \(\Kconst\), represented by
\begin{equation}
  \label{equ:kernel_function}
  \begin{aligned}
    \Kgauss(X_{i}, X_{j}, \Set{\theta_{1}, \theta_{2}})
    &= \theta_{1}^{2}\exp\left(-\frac{(X_{i} - X_{j})^{2}}{2\theta_{2}^{2}}\right)\\
    \Kconst(X_{i}, X_{j}, \Set{\theta_{3}})
    &= \theta_{3}^{2}\\
    K(X_{i}, X_{j}, \bm{\theta})
    &= \Kgauss(X_{i}, X_{j}, \Set{\theta_{1}, \theta_{2}}) + \Kconst(X_{i}, X_{j}, \Set{\theta_{3}}),\\
  \end{aligned}
\end{equation}
where \(\theta_{1}, \theta_{2}\), and \(\theta_{3}\) are hyperparameters.
This kernel is generally used in Gaussian process regression.
The Gaussian kernel function ensures smoothness and locality, whereas the constant kernel contributes to the global behavior.
(Although we initially used the polynomial kernel function under the assumption that local exponents are monotonic, we realized that they are not monotonic after applying this approach.)
For the stability of the regression, we convert the obtained data as
\begin{equation}
  \label{equ:converted-data}
  \begin{aligned}
    X_{i}
    &\equiv \frac{1}{\log(t_{i}) + c_{x}}\\
    Y_{i}
    &\equiv \frac{1}{c_{y1}\log(y_{i}) + c_{y2}}\\
    E_{i}
    &\equiv \frac{c_{y1}e_{y_{i}}}{y_{i}\left(c_{y1}\log(y_{i}) + c_{y2}\right)^{2}}\\
    c_{x}
    &\equiv 1 - \min_{i}\Set{\log(t_{i})}\\
    c_{y1}
    &\equiv \frac{\max_{i}\Set{\log(t_{i})} - \min_{i}\Set{\log(t_{i})}}{\max_{i}\Set{\log(y_{i})} - \min_{i}\Set{\log(y_{i})}}\\
    c_{y2}
    &\equiv 1 - c_{y1}\min_{i}\Set{\log(y_{i})},\\
  \end{aligned}
\end{equation}
where \(e_{i}\) represents the error in \(y_{i}\) and \(c_{x}\), \(c_{y1}\), and \(c_{y2}\) are constants determined by the data.
Note that the condition \(0 \le X_{i}, Y_{i} \le 1\) is satisfied, corresponding to normalization in machine learning.
This conversion remains invariant even if \(y_{i}\) is multiplied by a positive constant value, and it makes \(X_{i}\) and \(Y_{i}\) dimensionless quantities.
In the thermodynamic limit, \(y \to \infty\) as \(t \to \infty\) must be observed at the critical temperature for each case of \(y = 1/m, \chi, \mprime\).
Therefore, we include the data point \((X_{i}, Y_{i}, E_{i}) = (0, 0, 0)\) for the regressions of each case.
The local exponent at \(X = 1 / (\log(t) + c_{x})\) is represented by \cref{equ:prediction,equ:prediction-of-derivative} as
\begin{equation}
  \label{equ:prediction-of-lambda}
  \begin{aligned}
    \lambda_{y}(t)
    & = \frac{\partial \log(y)}{\partial \log(t)}\\
    & = \frac{\partial \log(y)}{\partial Y}\frac{\partial X}{\partial \log(t)}\frac{\partial Y}{\partial X}\\
    & = \frac{X^{2}}{c_{y1}Y^{2}}\frac{\partial Y}{\partial X}\\
    & = \frac{X^{2}}{c_{y1}\left(\ypredict(X)\right)^{2}}\frac{\partial \ypredict(X)}{\partial X}\\
    \frac{\partial X}{\partial \log (t)}
    &= - \frac{1}{(\log(t) + c_{x})^{2}} = - X^{2}\\
    \frac{\partial Y}{\partial \log (y)}
    &= - \frac{c_{y1}}{(c_{y1}\log(y) + c_{y2})^{2}} = - c_{y1}Y^{2}.\\
  \end{aligned}
\end{equation}
Note that, in contrast to the numerical differentiation, the present method enables us to obtain \(Y\) and \(\displaystyle \frac{\partial Y}{\partial X}\) at \(t\) as \(\ypredict\) and \(\displaystyle \frac{\partial \ypredict}{\partial X}\), respectively.
Therefore, we can predict local exponents \(\beta(t)\), \(\gamma(t)\), \(\nu(t)\), and \(z(t)\) from \cref{equ:critical_exponents,equ:prediction-of-lambda}.
Although we might be able to compute the exponent \(\alpha(t)\) directly from the relaxation of specific heat,~\cite{ozeki2007nonequilibrium} we did not calculate it in the present work because of its slow convergence.
Estimating \(\alpha(t)\) with the same accuracy as other critical exponents in the same observation time has long been considered difficult.
Of course, we can calculate \(\alpha\) using the scaling relation.

\subsection{Demonstration for the two-dimensional Ising model}
\begin{figure}[H]
  \centering
  \includegraphics[width=8.6cm]{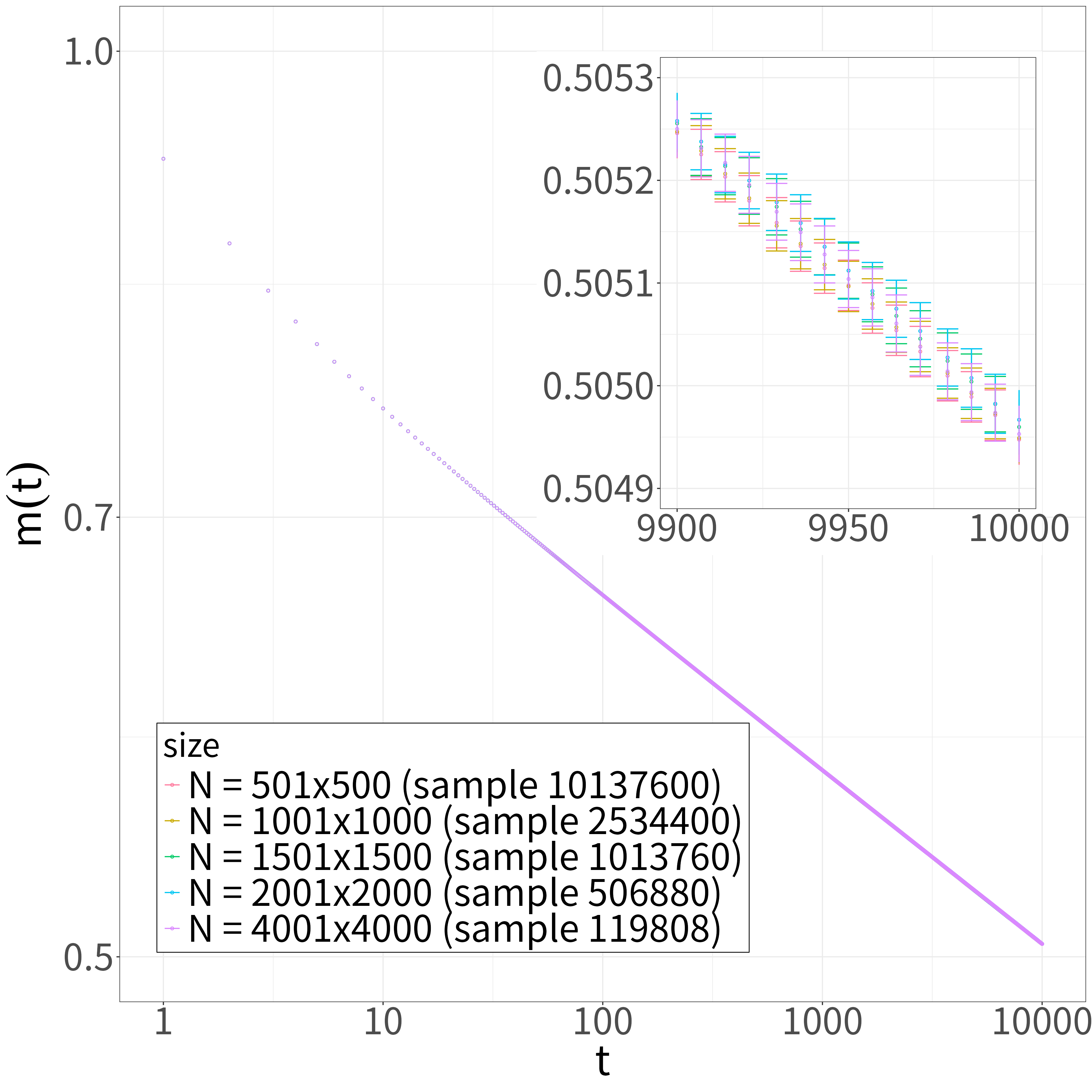}
  \caption{\label{fig:size-dependence-2D-Ising}
    (Color Online) Size dependence of the relaxation of \(m(t)\) from the all-up state for the two-dimensional Ising model, plotted on a double-logarithmic scale for \(1 \le t \le 10,000\) at the critical temperature \(T = 2.26918531421 = \Tc\).
    The inset gives \(m(t)\) between \(t = 9900\) and \(10,000\).
    The overlapping error bars for each lattice size suggest negligible size dependence for the \(\Nsize = 501 \times 500\) lattice up to \(10^{4}\) MCSs.
  }
\end{figure}

\begin{figure}[H]
  \centering
  \includegraphics[width=8.6cm]{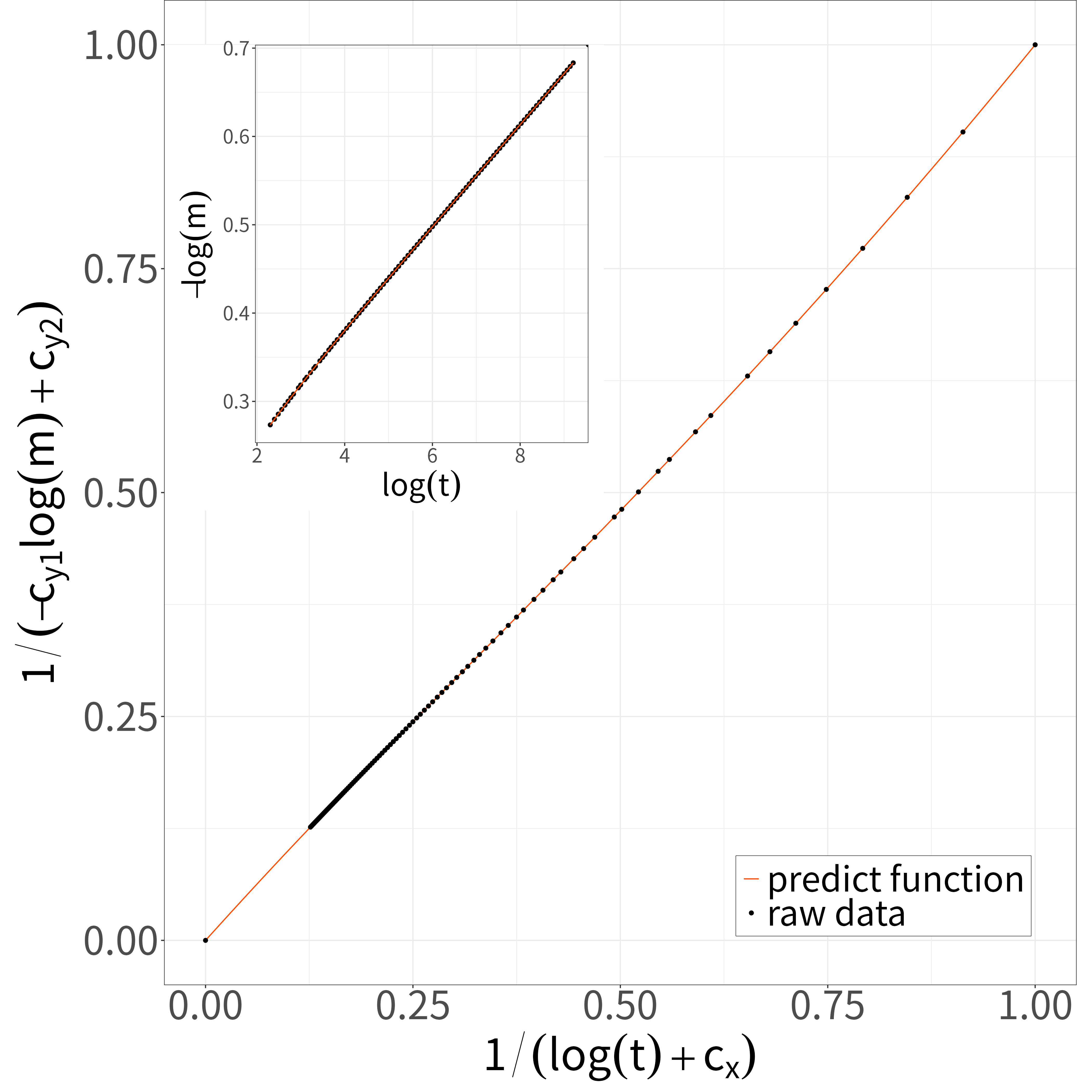}
  \caption{\label{fig:regression-scaled-2D-Ising-magne}
    (Color Online) Regression of \(X = 1 / (\log(t) + c_{x})\) vs \(Y = 1 / (-c_{y1}\log(m) + c_{y2})\) for the two-dimensional Ising model, within the bounds \(0 \le X, Y \le 1\).
    Black closed circles represent raw data points from the simulation, and the orange line depicts the interpolation.
    The inset shows the raw data of \(\log(t)\) vs \(-\log(m)\) and their prediction.
  }
\end{figure}

\begin{figure}[H]
  \centering
  \includegraphics[width=8.6cm]{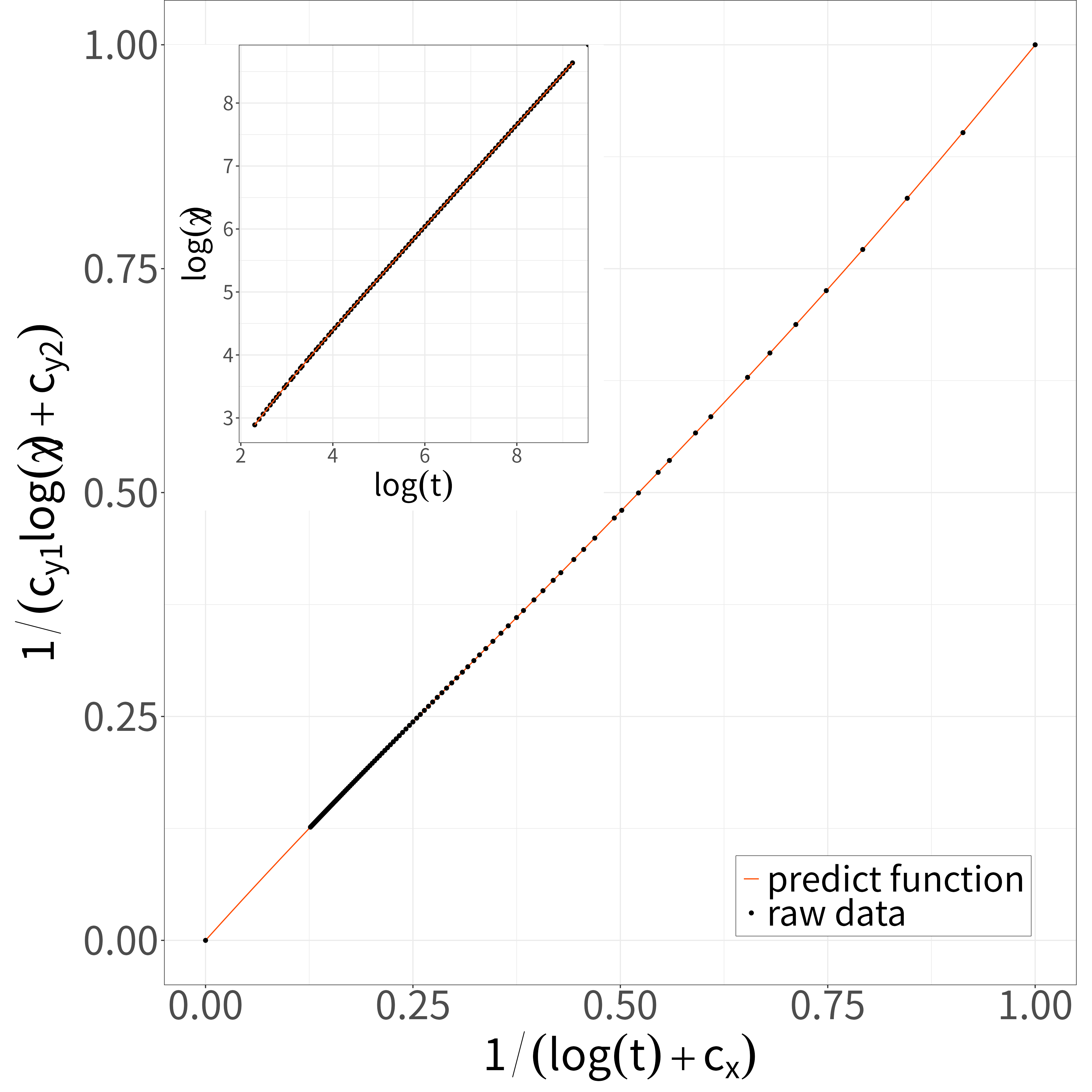}
  \caption{\label{fig:regression-scaled-2D-Ising-chi}
    (Color Online) Regression of \(X = 1 / (\log(t) + c_{x})\) vs \(Y = 1 / (c_{y1}\log(\chi) + c_{y2})\) for the two-dimensional Ising model, within the bounds \(0 \le X, Y \le 1\).
    Black closed circles represent raw data points from the simulation, and the orange line depicts the interpolation.
    The inset shows the raw data of \(\log(t)\) vs \(\log(\chi)\) and their prediction.
  }
\end{figure}

\begin{figure}[H]
  \centering
  \includegraphics[width=8.6cm]{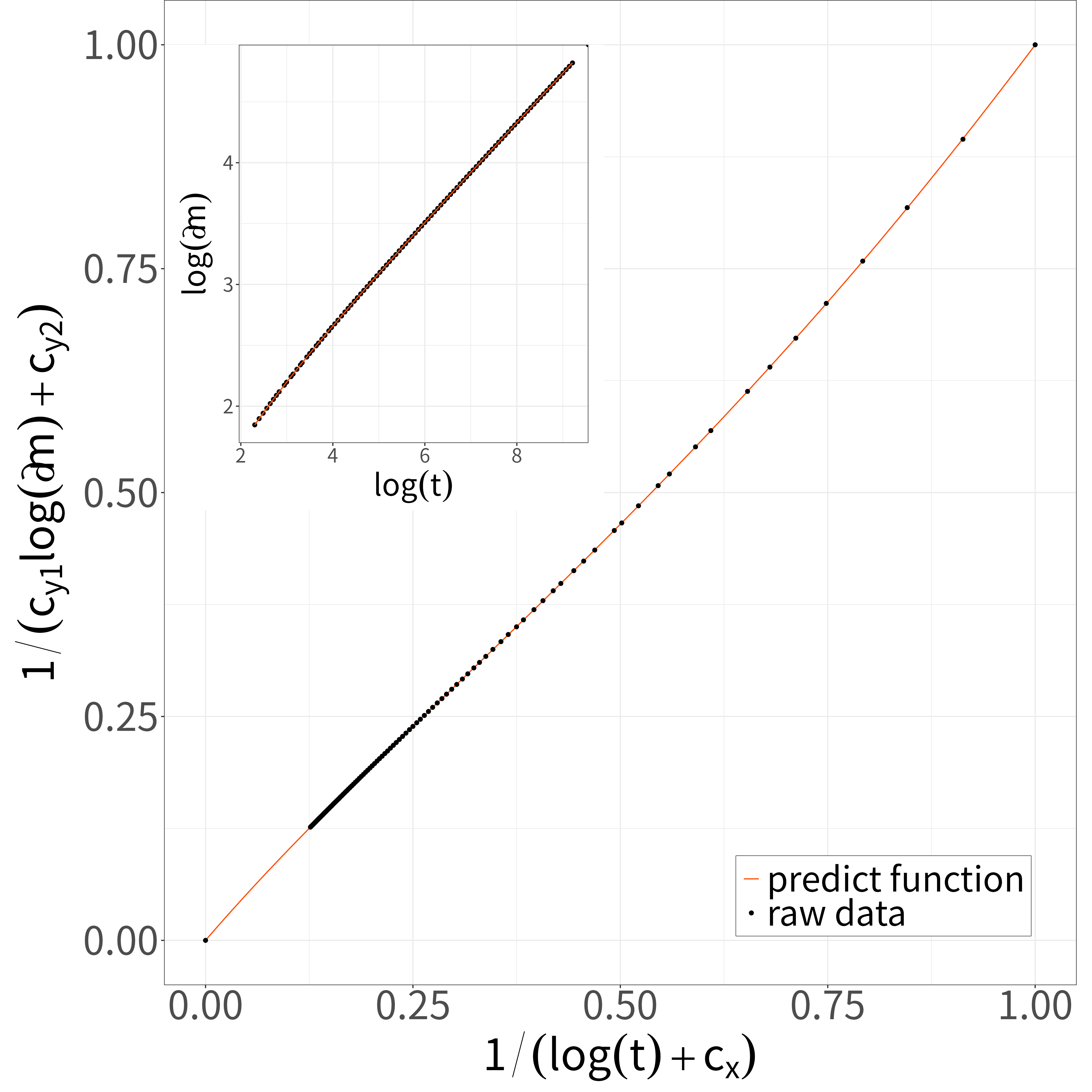}
  \caption{\label{fig:regression-scaled-2D-Ising-mprime}
    (Color Online) Regression of \(X = 1 / (\log(t) + c_{x})\) vs \(Y = 1 / (c_{y1}\log(\mprime) + c_{y2})\) for the two-dimensional Ising model, within the bounds \(0 \le X, Y \le 1\).
    Black closed circles represent raw data points from the simulation, and the orange line depicts the interpolation.
    The inset shows the raw data of \(\log(t)\) vs \(\log(\mprime)\) and their prediction.
  }
\end{figure}

Hereafter, measured temperatures are reported in units of \(J / \kb\).
We conducted simulations using the Metropolis algorithm on a \(\Nsize = 501 \times 500\) square lattice with skew boundary conditions at the critical temperature \(T = \Tc = 2.26918531421\).
An observation consists of \(10^{4}\) MCSs, with statistical averaging over \(10,137,600\) independent samples.
Initially, in NER analysis, we examine size dependence.
The overlapping error bars shown in \cref{fig:size-dependence-2D-Ising} indicate that the finite-size effect is negligible on a \(\Nsize = 501 \times 500\) lattice up to \(10^{4}\) MCSs.
Thus, we applied the present method using simulations with the \(\Nsize = 501 \times 500\) lattice to estimate local exponents.
For the regression analysis, we extracted \(\Ndata = 100\) data points at equal intervals of \(\log(t)\) from the simulation data, ranging from \(t = 10\) to \(t = 10,000\).
The results of the Gaussian process regression based on \cref{equ:converted-data} are shown in \cref{fig:regression-scaled-2D-Ising-magne,fig:regression-scaled-2D-Ising-chi,fig:regression-scaled-2D-Ising-mprime}.

\begin{figure}[H]
  \centering
  \includegraphics[width=8.6cm]{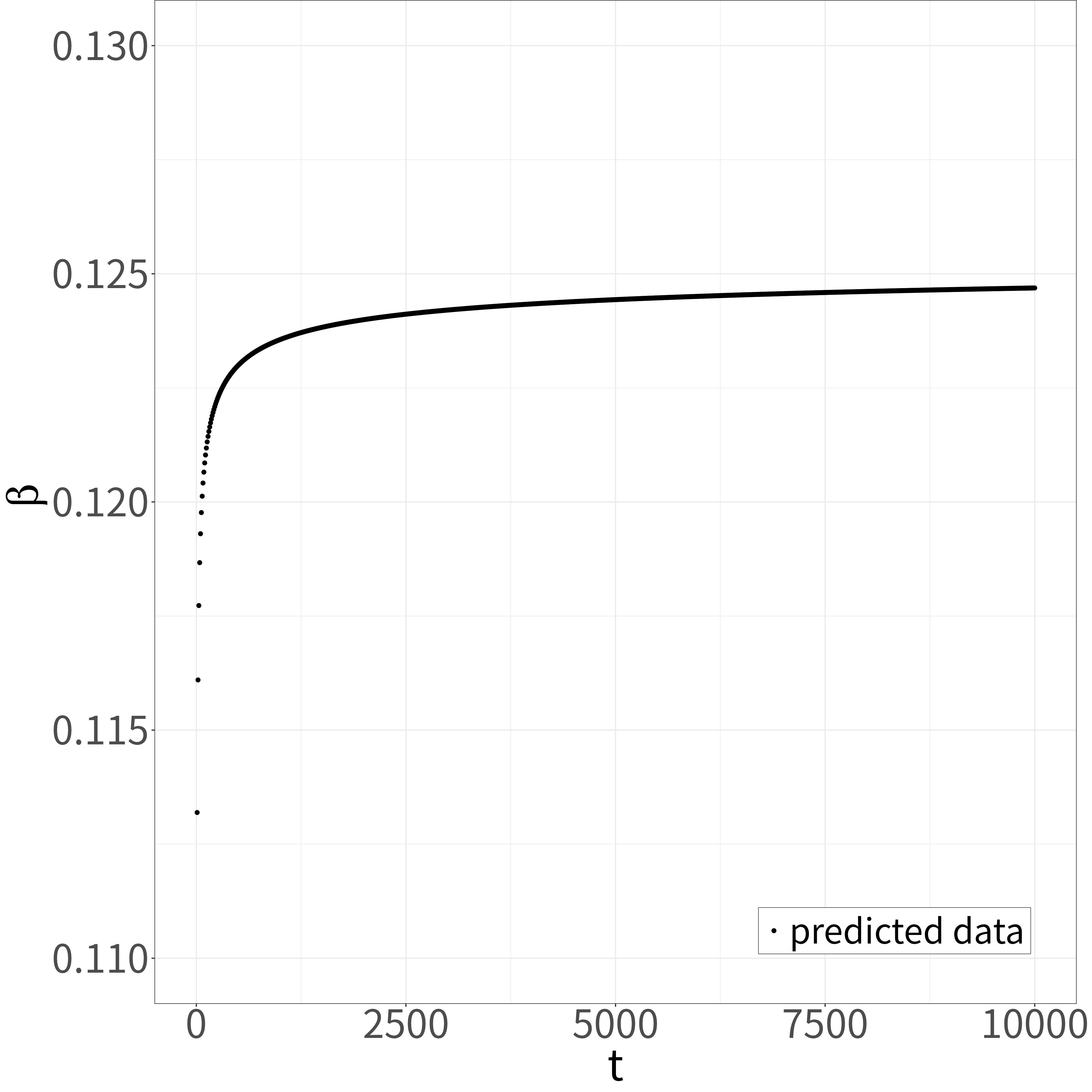}
  \caption{\label{fig:interpolation-beta-2D-Ising}
    Interpolation of \(t\) vs \(\beta\) for the two-dimensional Ising model.
    Black closed circles represent interpolated values at equal intervals of \(t\).
  }
\end{figure}

\begin{figure}[H]
  \centering
  \includegraphics[width=8.6cm]{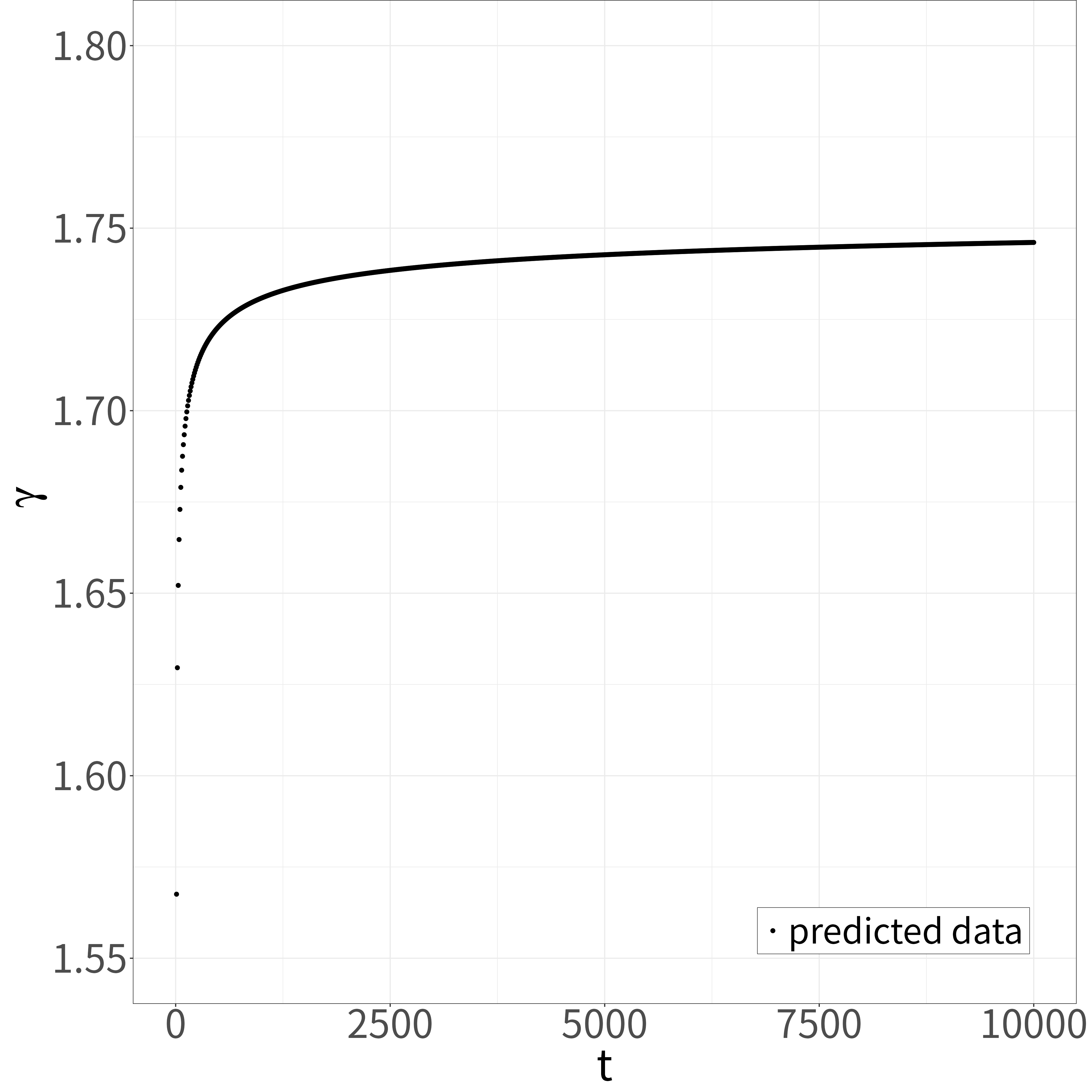}
  \caption{\label{fig:interpolation-gamma-2D-Ising}
    Interpolation of \(t\) vs \(\gamma\) for the two-dimensional Ising model.
    Black closed circles represent interpolated values at equal intervals of \(t\).
  }
\end{figure}

\begin{figure}[H]
  \centering
  \includegraphics[width=8.6cm]{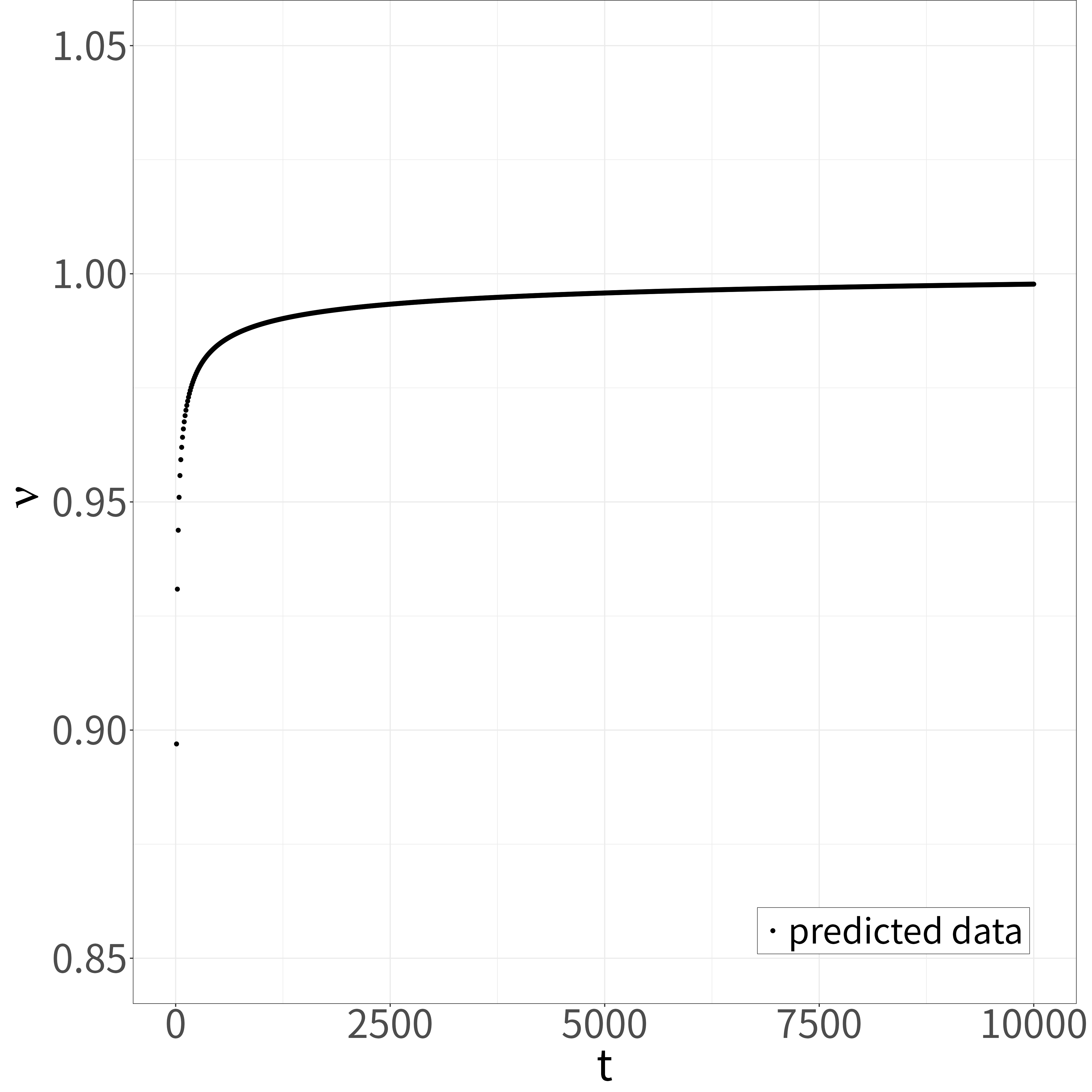}
  \caption{\label{fig:interpolation-nu-2D-Ising}
    Interpolation of \(t\) vs \(\nu\) for the two-dimensional Ising model.
    Black closed circles represent interpolated values at equal intervals of \(t\).
  }
\end{figure}

\begin{figure}[H]
  \centering
  \includegraphics[width=8.6cm]{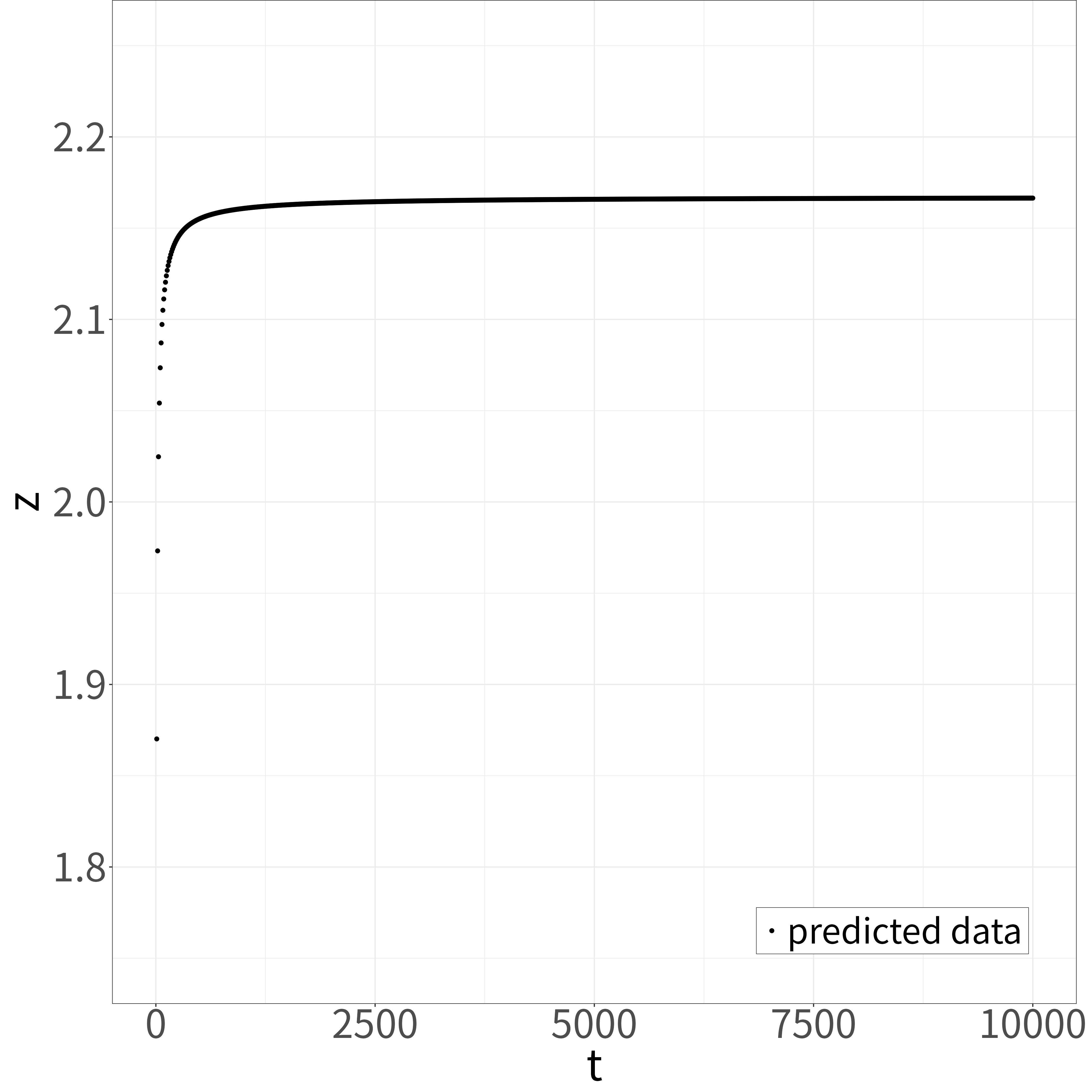}
  \caption{\label{fig:interpolation-z-2D-Ising}
    Interpolation of \(t\) vs \(z\) for the two-dimensional Ising model.
    Black closed circles represent interpolated values at equal intervals of \(t\).
  }
\end{figure}

\Cref{fig:interpolation-beta-2D-Ising,fig:interpolation-gamma-2D-Ising,fig:interpolation-nu-2D-Ising,fig:interpolation-z-2D-Ising} display the local exponents in the observed time interval, as estimated using \cref{equ:critical_exponents,equ:prediction-of-lambda}.
We plotted \(N = 1003\) data points at equal intervals of \(t\) in these figures.
In contrast to numerical differentiation, our approach predicts derivatives at specific times.
Note that we predict and use values in the observed time interval \(10 \le t \le 10,000\) because predictions outside this interval are unstable.
If we use an interval around \((X, Y) = (0, 0)\) for the interpolation by regression, the value of \(\displaystyle \frac{X^{2}}{\ypredict^{2}}\) in \cref{equ:prediction-of-lambda} close to that interval is also unstable because of the small values in the fraction.
\begin{figure}[H]
  \centering
  \includegraphics[width=8.6cm]{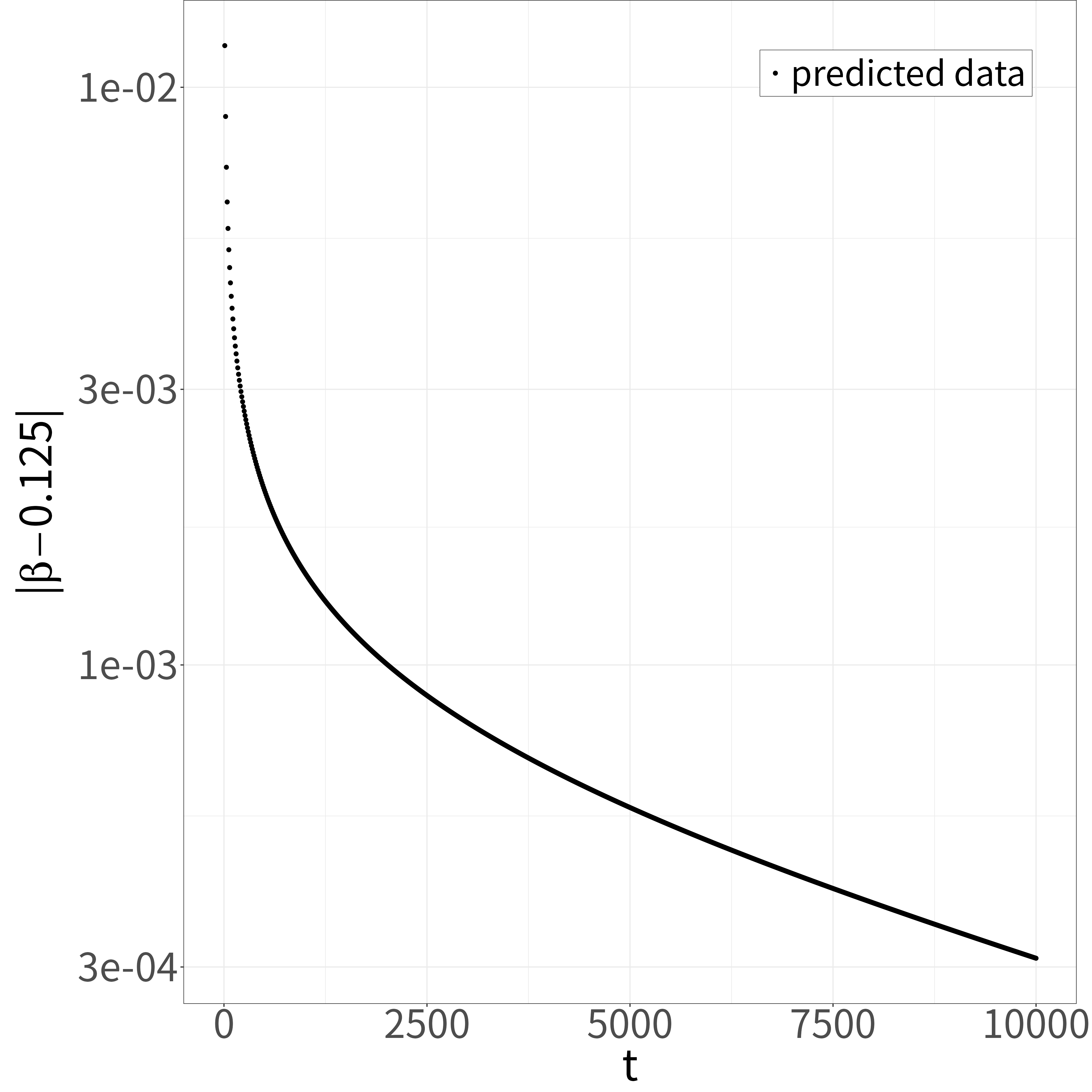}
  \caption{\label{fig:interpolation-diff_beta-2D-Ising}
    Interpolation of \(t\) vs \(\abs{\beta - 0.125}\) for the two-dimensional Ising model, plotted on a semilogarithmic graph with the y-axis on a logarithmic scale.
    Black closed circles represent interpolated values at equal intervals of \(t\).
  }
\end{figure}

\begin{figure}
    \centering
    \includegraphics[width=8.6cm]{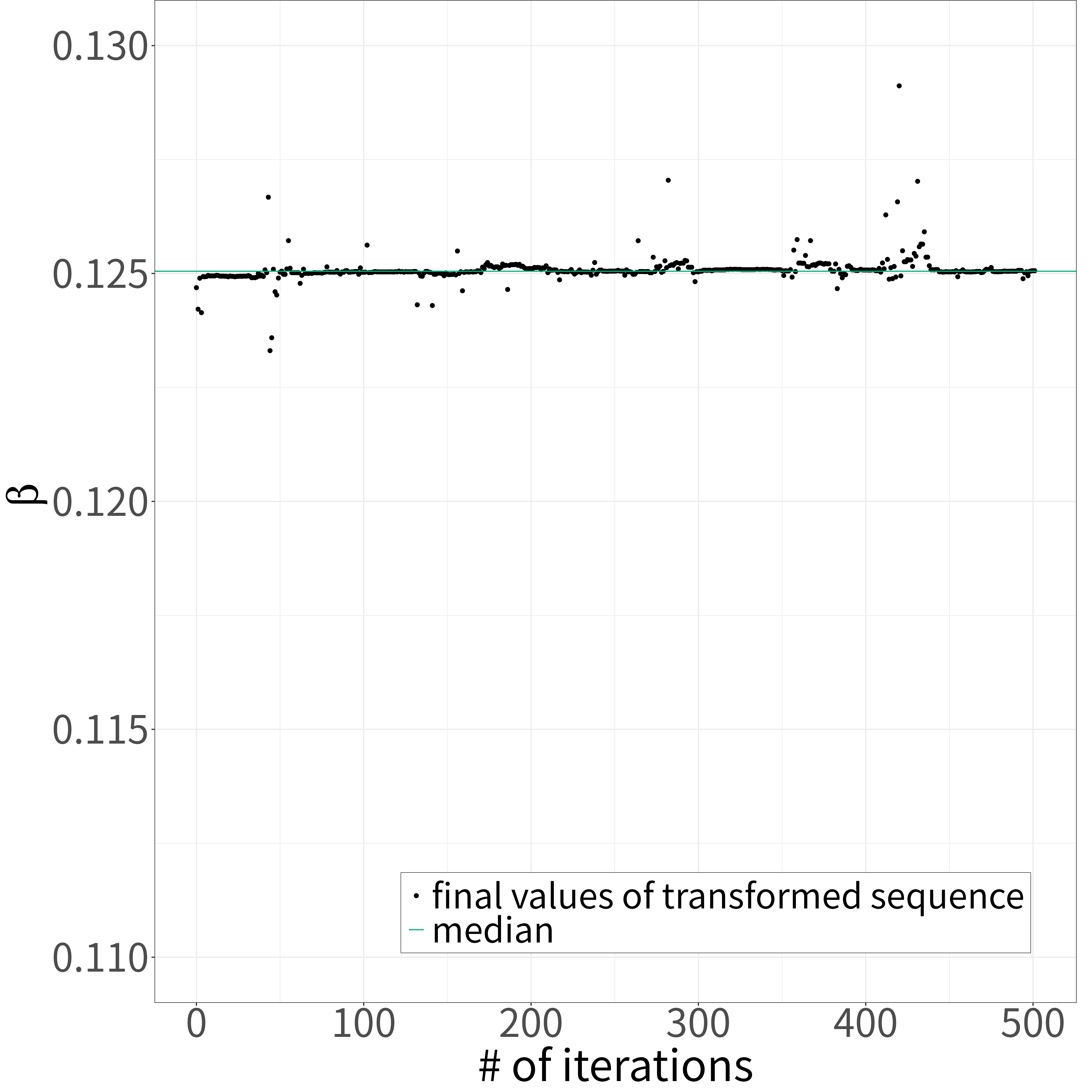}
    \caption{\label{fig:extrapolation-beta-2D-Ising}
      (Color Online) Extrapolation of the number of iterations \(k\) vs \(\beta\) for the two-dimensional Ising model with the length of the original sequence \(N = 1003\).
      Black closed circles represent the final values of the transformed sequence.
      The green line represents their median.
    }
\end{figure}

Finally, we apply the extrapolation method for systematicity and reproducibility to estimate critical exponents as \(t \to \infty\).
We use the \(\varepsilon\)-algorithm,~\cite{wynn1956device,sidi2003practical,brezinski2020extrapolation} which is an efficient implementation of the Shanks transformation.~\cite{shanks1955non}
Briefly, the Shanks transformation extrapolates sequences that converge exponentially.
In \cref{fig:interpolation-diff_beta-2D-Ising},  we replotted \(\beta(t)\) shown in \cref{fig:interpolation-beta-2D-Ising} in order to see the exponential convergence in the asymptotic region.
Thus we assume that local exponents such as \(\beta(t)\) converge exponentially with respect to \(t\); hereafter, we apply the \(\varepsilon\)-algorithm to the sequences obtained from regressions.
Now we define the sequence \(v_{s}\) as
\begin{equation}
  \label{equ:function-to-sequence}
  \begin{aligned}
    v_{s}
    &\equiv v\left(\frac{(N - 1 - s)\tmin + s\tmax}{N - 1}\right), \quad s = 0, \ldots, N - 1\\
  \end{aligned}
\end{equation}
where \(s\) is the label for data, \(\tmin\)(\(\tmax\)) is the minimum (maximum) time in the interval, \(N\) denotes the number of data, and \(v(t)\) represents \(\beta(t)\), \(\gamma(t)\), \(\nu(t)\), or \(z(t)\).
The \(\varepsilon\)-algorithm progresses as follows:
\begin{equation}
  \label{equ:epsilon-table}
    \frac{1}{\varepsilon_{s - 1, k + 1} - \varepsilon_{s, k}} + \frac{1}{\varepsilon_{s + 1, k - 1} - \varepsilon_{s, k}}
    = \frac{1}{\varepsilon_{s - 1, k} - \varepsilon_{s, k}} + \frac{1}{\varepsilon_{s + 1, k} - \varepsilon_{s, k}}, \quad s, k = 0, 1, \ldots,
\end{equation}
starting with \(\varepsilon_{s, -1} = \infty\), and \(\varepsilon_{s, 0} = v_{s}\) for \(s = 0, \ldots, N - 1\).
We can obtain the convergence-accelerated sequence \(\varepsilon_{s, k}\) by applying the transformation \(k\) times, where the length is \(N - 2k\).
In the present study, we estimate the critical exponent by taking the median of the final values in the extrapolated sequence \(\varepsilon_{s, k}\) for \(k = 0, \ldots, \floor{(N - 1)/2}\), with \(N = 1003\) data points at equal intervals of \(t\).
Specifically, we calculate the limit \(V\) using
\begin{equation}
  \label{equ:median-of-final-values}
  V = \median\left({\set{\vfinal \mid \vfinal = \varepsilon_{N - 1 - 2k, k}, k = 0, \ldots, \floor{(N-1)/2}}}\right)
\end{equation}
as the estimator of the critical exponent.
Note that we opted to use the median and \(N = 1003\) interpolated data to mitigate the effect of outliers given that the actual data sequence may not perfectly adhere to \cref{equ:epsilon-table} because of numerical errors.
Because most of the final values shown in \cref{fig:extrapolation-beta-2D-Ising} are close to the median, we regard the median as the limit to neglect outliers.
We obtain \(\beta = 0.12507\ldots\), \(\gamma = 1.7508\ldots\), \(\nu = 1.0004\ldots\), and \(z = 2.1668\ldots\), which closely approximate the exact values \(\beta = 0.125\), \(\gamma = 1.75\), and \(\nu = 1\) and are consistent with previously reported results \(z = 2.1667(5)\).~\cite{nightingale2000monte}

Because the present method operates automatically, reproducibly, and reliably, we can easily apply the bootstrap method.~\cite{efron1979bootstrap}
We created \(100\) bootstrap samples by resampling \(\Ndata = 100\) data points, extracted at equal intervals of \(\log(t)\), \(100\) times.
We independently applied the present method to each bootstrap sample and estimated the mean and numerical errors of the critical exponents across the bootstrap samples.
Consequently, we estimated the critical exponents as \(\beta = 0.12504(6)\), \(\gamma = 1.7505(10)\), and \(\nu = 1.0003(6)\), which are consistent with the exact values \(\beta = 0.125\), \(\gamma = 1.75\), and \(\nu = 1\).
The value \(z = 2.1669(9)\) is reliable because of the high accuracy of these exponents.
Our estimation of the dynamical exponent \(z = 2.1669(9)\) is consistent with that reported in a previous study [\(z = 2.1667(5)\)]~\cite{nightingale2000monte} and is close to that reported in another study [\(z = 2.14(2)\)].~\cite{adzhemyan2022dynamic}

We also applied the present method to the same model over various time intervals.
The results are shown in \cref{tab:compare-2D-Ising}.
As the upper time interval increases, the estimation accuracy for critical exponents improves and the exponents become consistent with the exact values and with the results of the previous study.~\cite{nightingale2000monte}
These results validate the accuracy and reliability of the present method at the critical temperature and demonstrate the sufficiency of the interval \(t = [10, 10,000]\).
Data from a simple simulation for the relaxation of the appropriate quantities enable us to derive highly reliable critical exponents.
\begin{widetext}
  \begin{table*}
    \caption{\label{tab:compare-2D-Ising}
    Critical exponents for the two-dimensional Ising model from various studies.
    }
    \begin{tabular}{lclllll}
      \hline\hline
      Reference & Method & \(\beta\) & \(\gamma\) & \(\nu\) & \(z\)\\
      \hline
      Exact values & & \(0.125\) & \(1.75\) & \(1\) & \\
      Nightingale \emph{et al.} (2000)~\cite{nightingale2000monte} & MC & & & & \(2.1667(5)\) \\
      Adzhemyan \emph{et al.} (2022)~\cite{adzhemyan2022dynamic} & \(\varepsilon\) expansion & & & & \(2.14(2)\) &\\
      Our results (\(10 \le t \le 1000\)) & NER & \(0.1241(1)\) & \(1.7364(14)\) & \(0.9923(8)\) & \(2.1719(5)\)\\
      Our results (\(10 \le t \le 5000\)) & NER & \(0.12473(6)\) & \(1.7438(9)\) & \(0.9966(5)\) & \(2.1715(9)\)\\
      Our results (\(10 \le t \le 10,000\)) & NER & \(0.12504(6)\) & \(1.7505(10)\) & \(1.0003(6)\) & \(2.1669(9)\)\\
      \hline\hline
    \end{tabular}
  \end{table*}
\end{widetext}

\subsection{Advantage over the conventional method}
\begin{figure}[H]
  \centering
  \includegraphics[width=8.6cm]{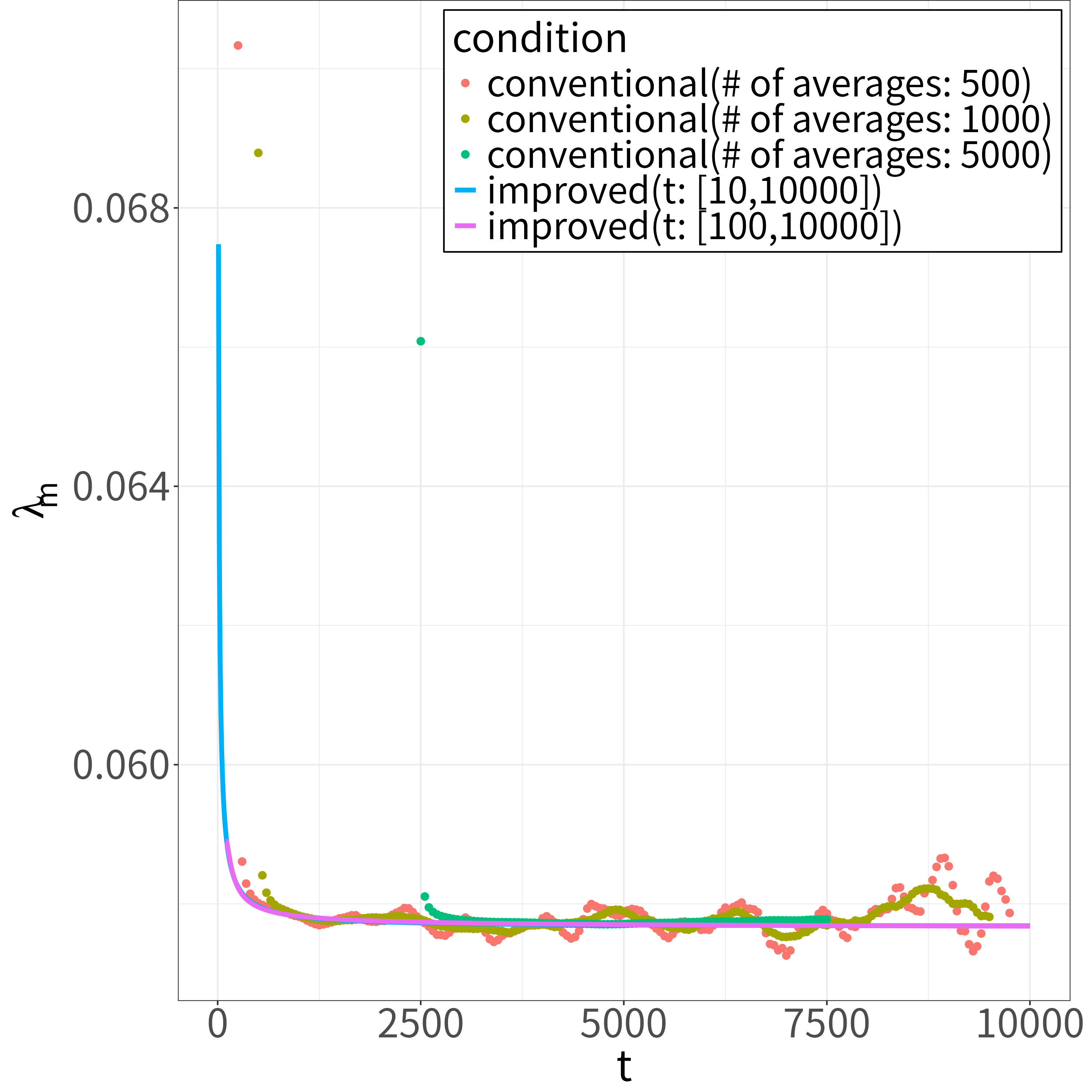}
  \caption{\label{fig:lambda_m-compare-2D-Ising}
    (Color Online) \(\lambda_{m}\) obtained using the conventional method and that obtained using the improved method.
    Circles represent the values obtained using the conventional method for several average numbers.
    Lines represent the values obtained using the improved method for several regression intervals.
    For the conventional method, the data points lie at equal intervals of \(t\) in increments of \(50\).
    }
\end{figure}

Because we have numerically demonstrated that the present method is reliable, we also illustrate the improvement visually.
In the conventional method, we compute \(\lambda_{m}\) from a linear approximation of sections.
The controllable conditions are the number of averages \(\Naverage\) and the choice of sections.
Therefore, we calculate \(\lambda_{m}\) at \(t = (t_{l} + t_{r}) / 2\) as
\[
  \lambda_{m}\left(\frac{t_{l} + t_{r}}{2}\right)
  = (\text{slope of \(-\log(m)\) over } [t_{l}, t_{r}]),
\]
where data points lie at equal intervals in \(t\), both \(t_{l}\) and \(t_{r}\) are integers, and \(t_{r} - t_{l} + 1 = \Naverage\).
In contrast to the conventional method, the controllable condition for the present method is the time interval for the regression.
\Cref{fig:lambda_m-compare-2D-Ising} shows a plot of \(\lambda_{m}\) values obtained under several controllable conditions using the conventional method as well as those obtained using the improved method.
The data points for both methods exhibit similar trends.
Although the data points for the conventional method are sensitive to the number of averages, those for the present method are less affected by the choice of the regression interval.
This plot indicates that the present method has improved reliability compared with the conventional method and overcomes discreteness.

\section{\NoCaseChange{Application to the Three-dimensional Ising Model}}
\label{sec:application-3D-ising}
\begin{figure}[H]
  \centering
  \includegraphics[width=8.6cm]{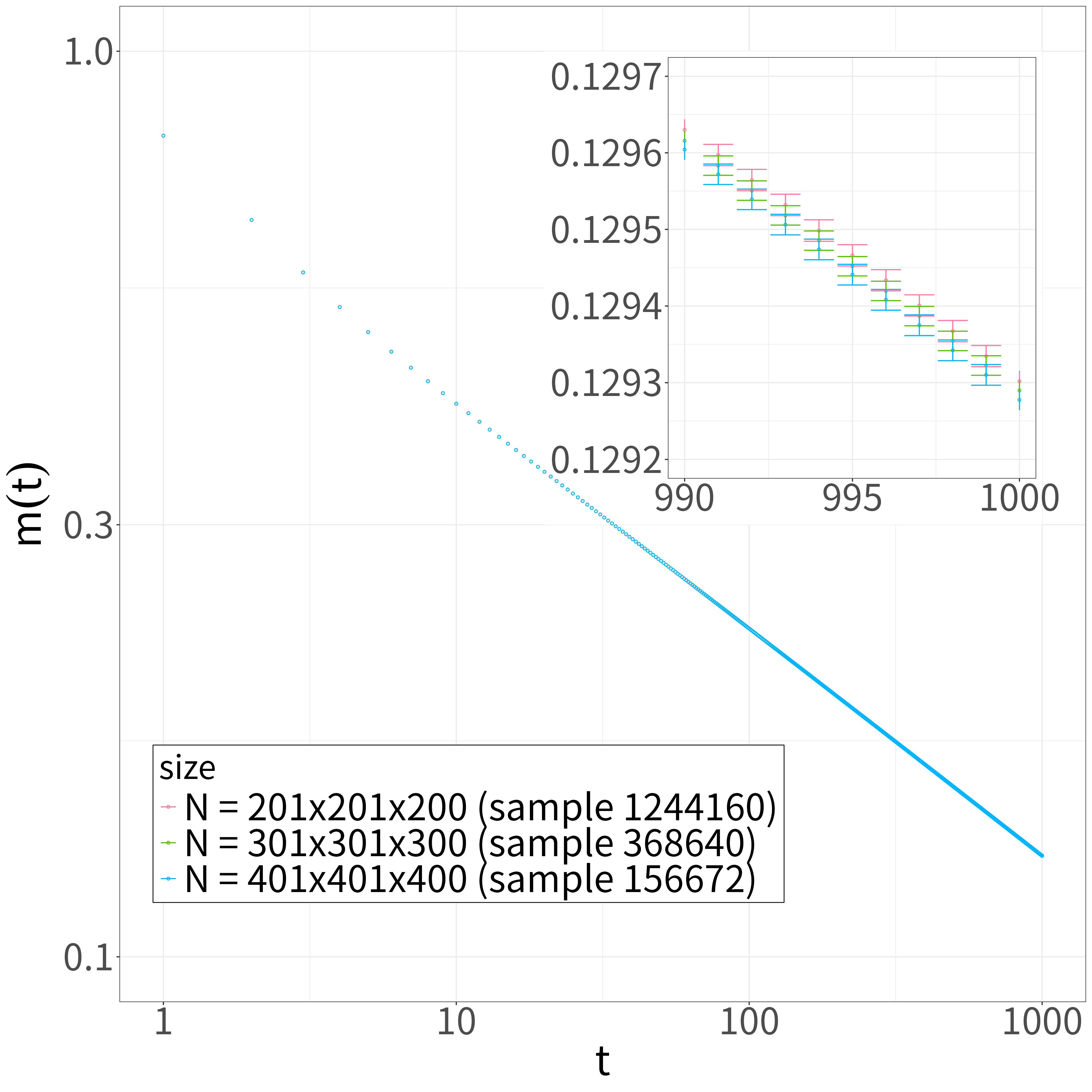}
  \caption{\label{fig:size-dependence-3D-Ising}
    (Color Online) Size dependence of the relaxation of \(m(t)\) from the all-up state for the three-dimensional Ising model, plotted on a double-logarithmic scale for \(1 \le t \le 1000\) at temperature \(T = 4.51152174982078 \simeq \Tc\).
    The inset illustrates \(m(t)\) between \(t = 990\) and \(1000\).
    The overlapping error bars for each lattice size suggest negligible size dependence for the \(\Nsize = 201 \times 201 \times 200\) lattice up to \(10^{3}\) MCSs.
  }
\end{figure}

\begin{figure}[H]
  \centering
  \includegraphics[width=8.6cm]{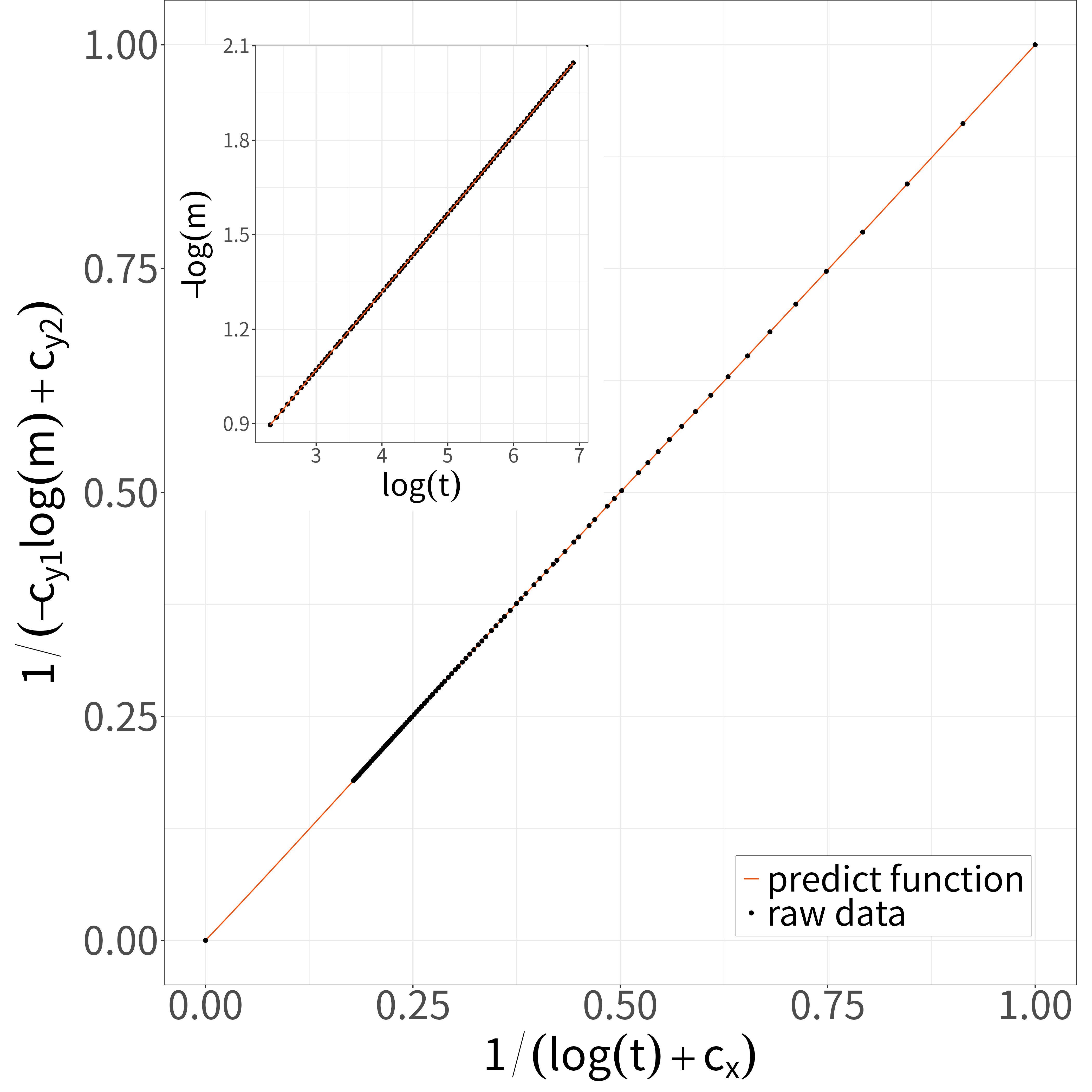}
  \caption{\label{fig:regression-scaled-3D-Ising-magne}
    (Color Online) Regression of \(X = 1 / (\log(t) + c_{x})\) vs \(Y = 1 / (-c_{y1}\log(m) + c_{y2})\) for the three-dimensional Ising model, within the bounds \(0 \le X, Y \le 1\).
    Black closed circles represent raw data points from the simulation; the orange line depicts the interpolation.
    The inset shows the raw data of \(\log(t)\) vs \(-\log(m)\) and their prediction.
  }
\end{figure}

\begin{figure}[H]
  \centering
  \includegraphics[width=8.6cm]{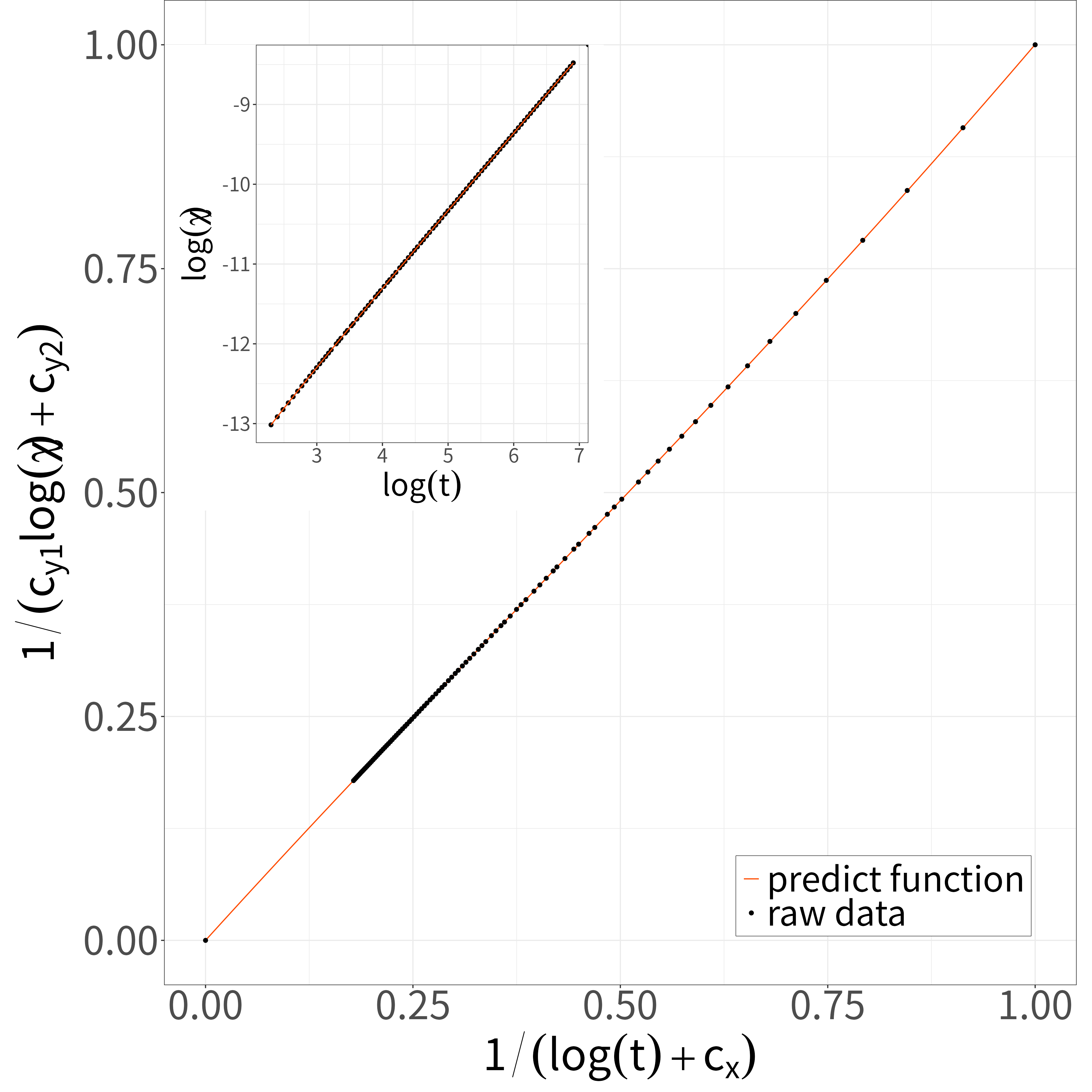}
  \caption{\label{fig:regression-scaled-3D-Ising-chi}
    (Color Online) Regression of \(X = 1 / (\log(t) + c_{x})\) vs \(Y = 1 / (c_{y1}\log(\chi) + c_{y2})\) for the three-dimensional Ising model, within the bounds \(0 \le X, Y \le 1\).
    Black closed circles represent raw data points from the simulation; the orange line depicts the interpolation.
    The inset shows the raw data of \(\log(t)\) vs \(\log(\chi)\) and their prediction.
  }
\end{figure}

\begin{figure}[H]
  \centering
  \includegraphics[width=8.6cm]{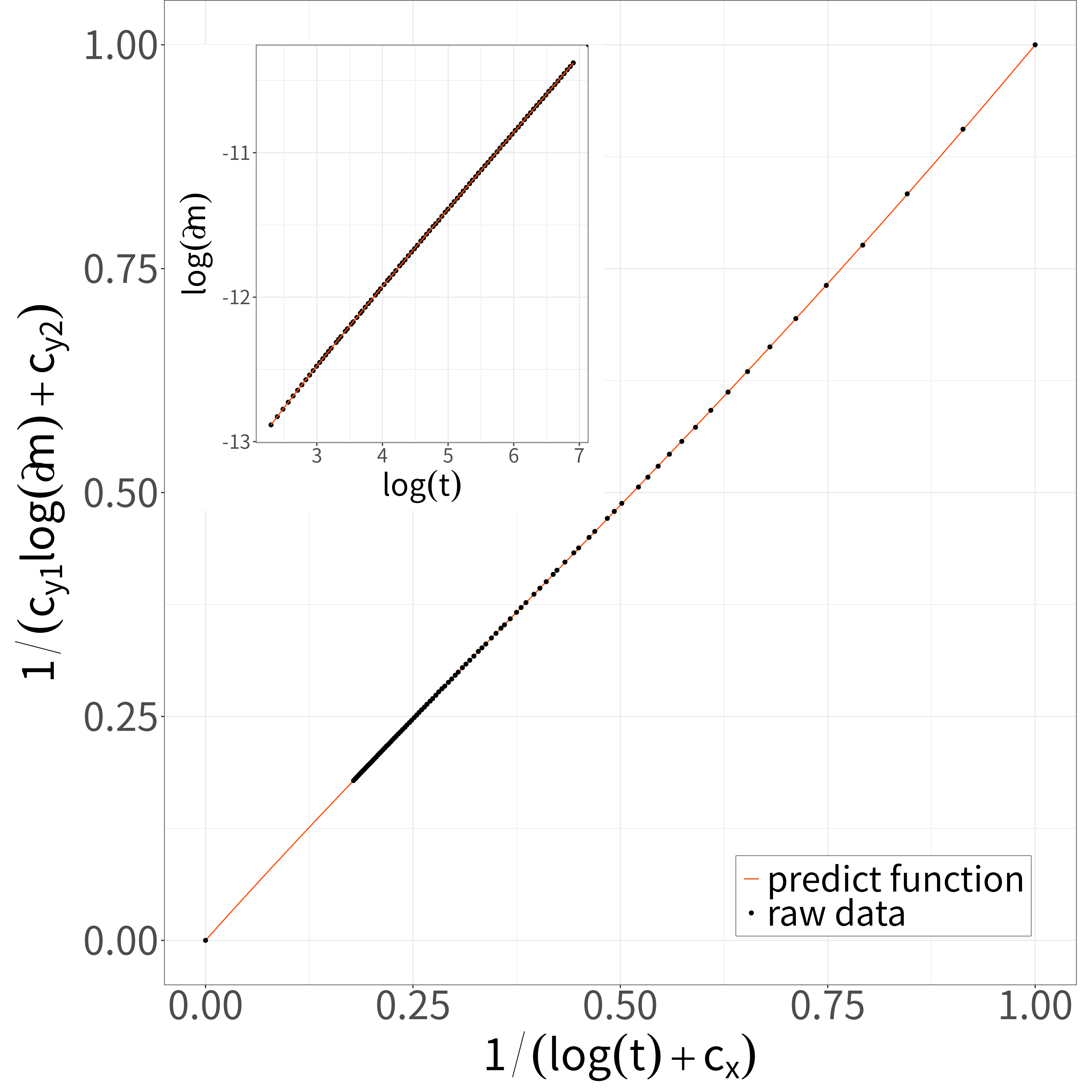}
  \caption{\label{fig:regression-scaled-3D-Ising-mprime}
    (Color Online) Regression of \(X = 1 / (\log(t) + c_{x})\) vs \(Y = 1 / (c_{y1}\log(\mprime) + c_{y2})\) for the three-dimensional Ising model, within the bounds \(0 \le X, Y \le 1\).
    Black closed circles represent raw data points from the simulation; the orange line depicts the interpolation.
    The inset shows the raw data of \(\log(t)\) vs \(\log(\mprime)\) and their prediction.
  }
\end{figure}

Because the analysis for the two-dimensional Ising model at the exact critical temperature was successful, we applied the proposed method to the three-dimensional cubic Ising model, whose exact transition temperature is unknown.
We conducted analyses of this model at the temperature \(T = 1 / 0.2216547 = 4.51152174982078\).
This temperature was estimated in a previous study using the pinching estimation of the NER method~\cite{ito2005nonequilibrium} explained in \cref{sec:NER-explanation}.
Simulations were performed on a \(\Nsize = 201 \times 201 \times 200\) cubic lattice with skew boundary conditions at \(T = 4.51152174982078\).
Observations consisted of \(10^{3}\) MCSs, with statistical averaging over \(1,244,160\) independent samples.
The overlapping error bars shown in \cref{fig:size-dependence-3D-Ising} indicate that the finite-size effect is negligible on a \(\Nsize = 201 \times 201 \times 200\) lattice up to \(10^{3}\) MCSs.
Similar to the above analysis, we applied the proposed method to the three-dimensional Ising model.
The regressions are shown in \cref{fig:regression-scaled-3D-Ising-magne,fig:regression-scaled-3D-Ising-chi,fig:regression-scaled-3D-Ising-mprime}.

\begin{figure}[H]
  \centering
  \includegraphics[width=8.6cm]{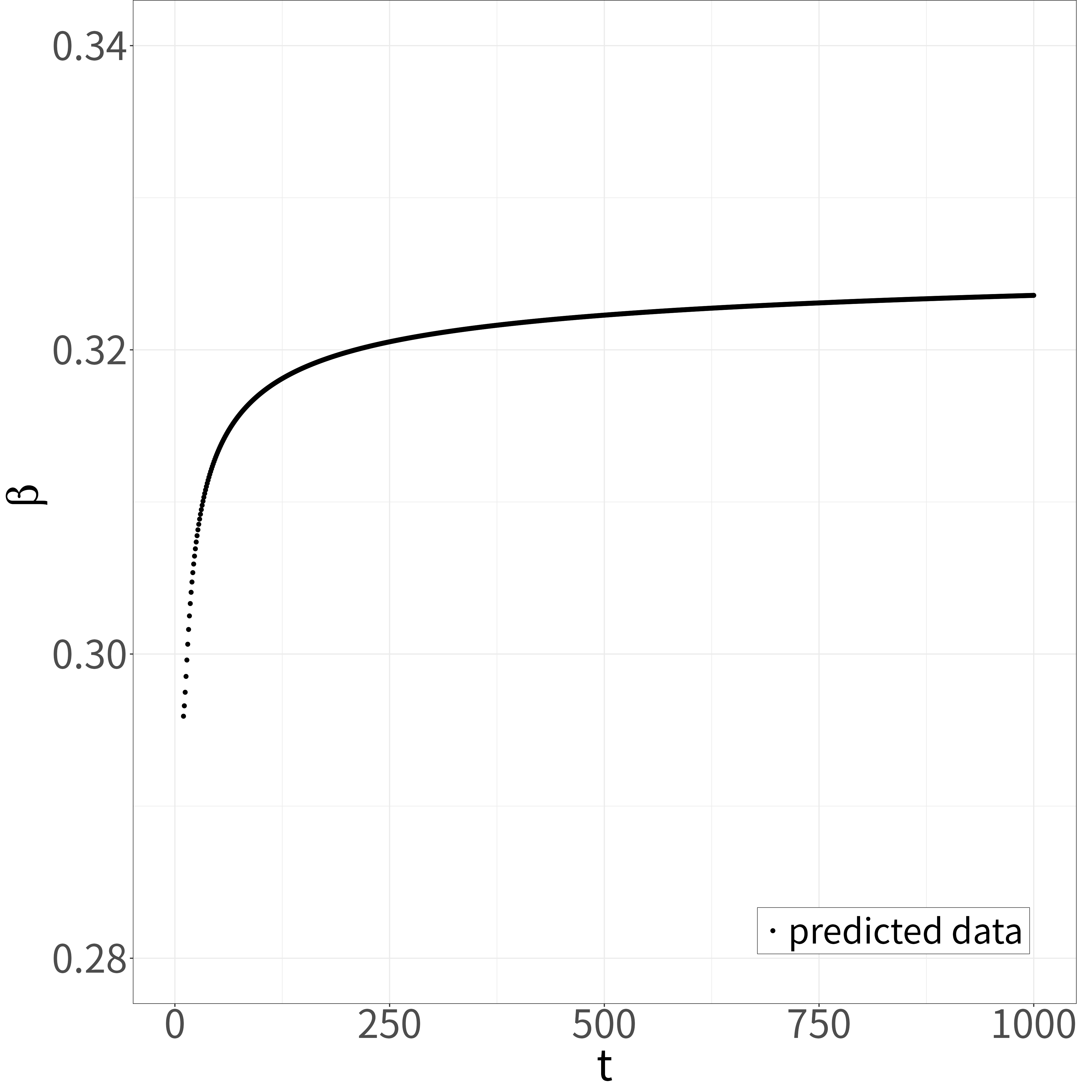}
  \caption{\label{fig:interpolation-beta-3D-Ising}
    Interpolation of \(t\) vs \(\beta\) for the three-dimensional Ising model.
    Black closed circles represent interpolated values at equal intervals of \(t\).
  }
\end{figure}

\begin{figure}[H]
  \centering
  \includegraphics[width=8.6cm]{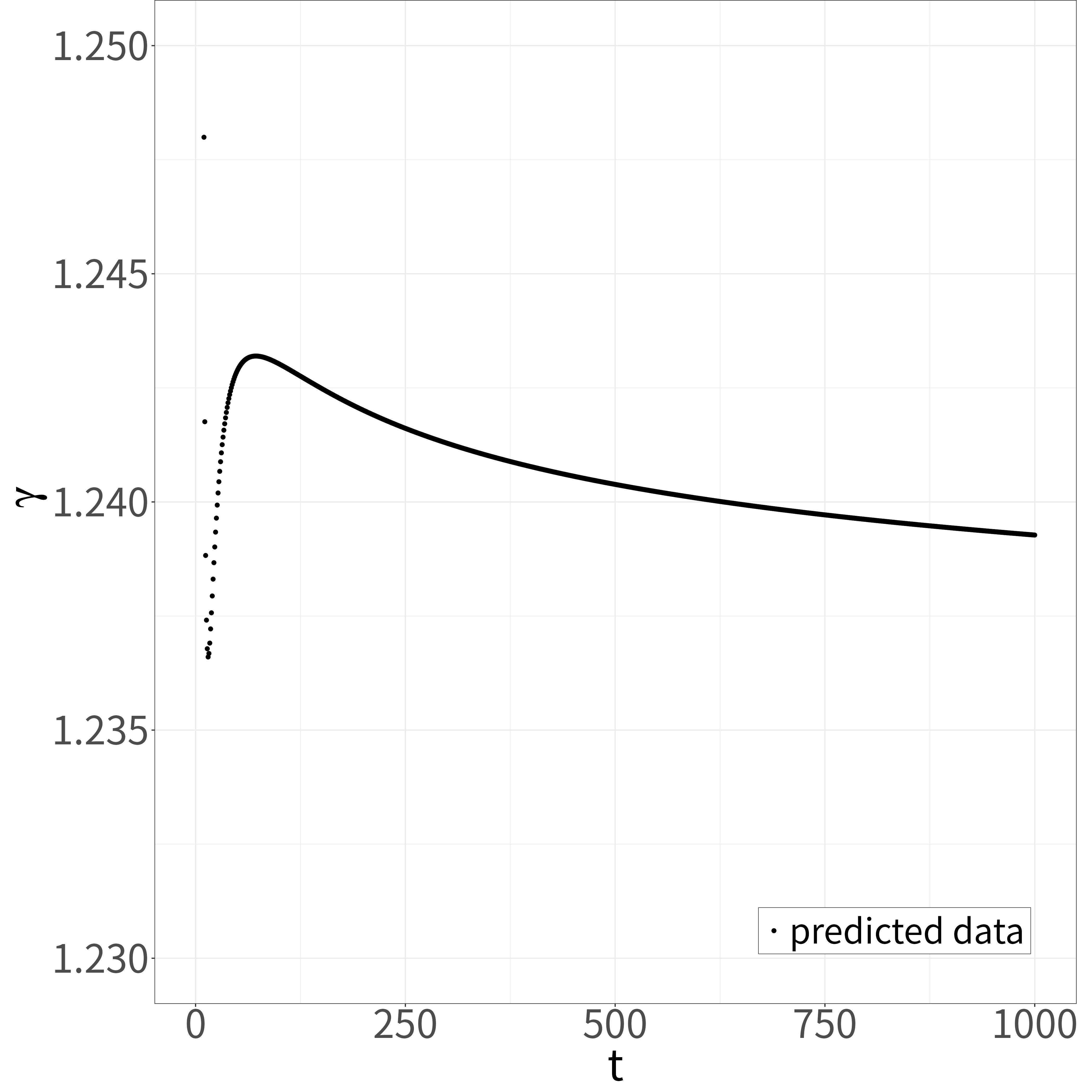}
  \caption{\label{fig:interpolation-gamma-3D-Ising}
    Interpolation of \(t\) vs \(\gamma\) for the three-dimensional Ising model.
    Black closed circles represent interpolated values at equal intervals of \(t\).
  }
\end{figure}

\begin{figure}[H]
  \centering
  \includegraphics[width=8.6cm]{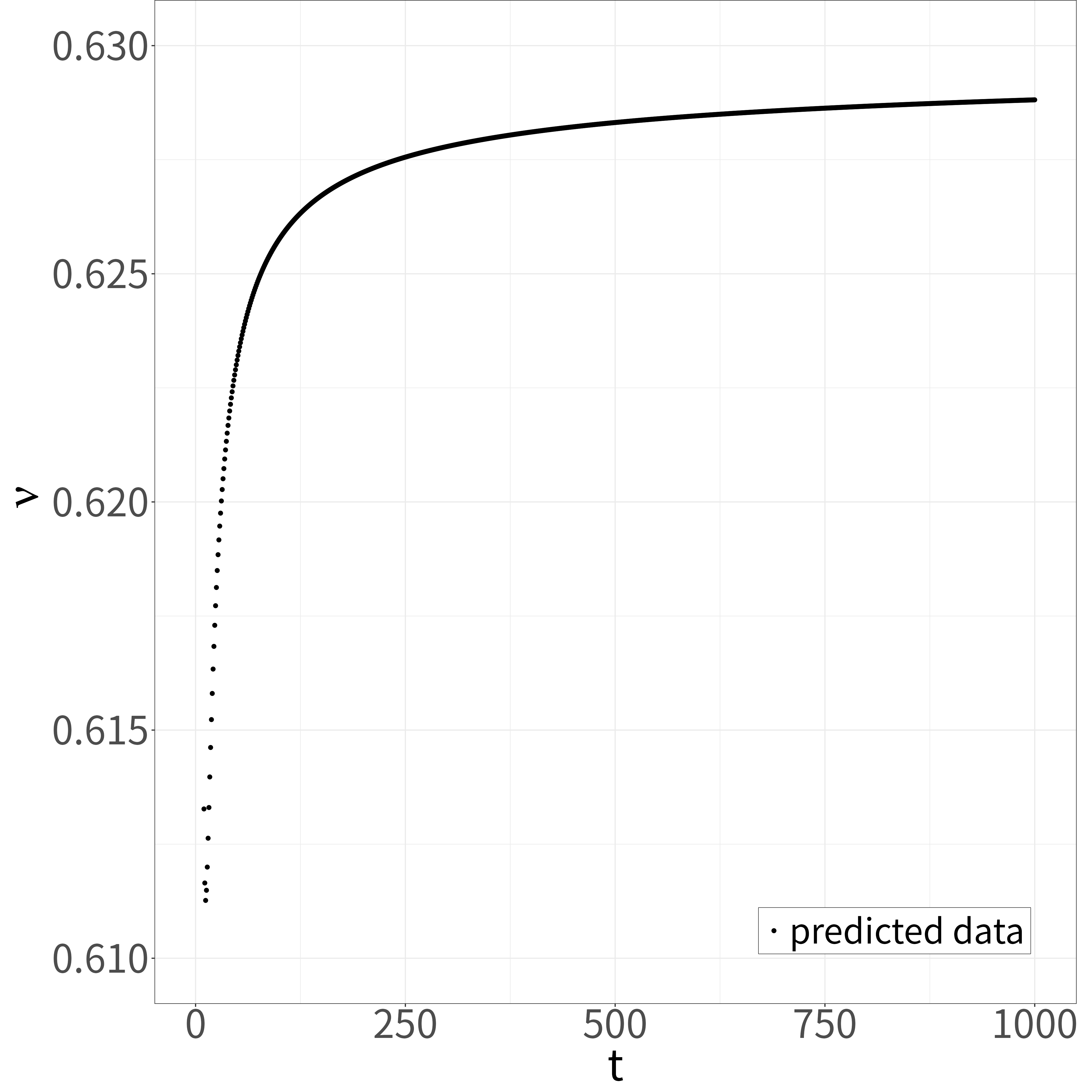}
  \caption{\label{fig:interpolation-nu-3D-Ising}
    Interpolation of \(t\) vs \(\nu\) for the three-dimensional Ising model.
    Black closed circles represent interpolated values at equal intervals of \(t\).
  }
\end{figure}

\begin{figure}[H]
  \centering
  \includegraphics[width=8.6cm]{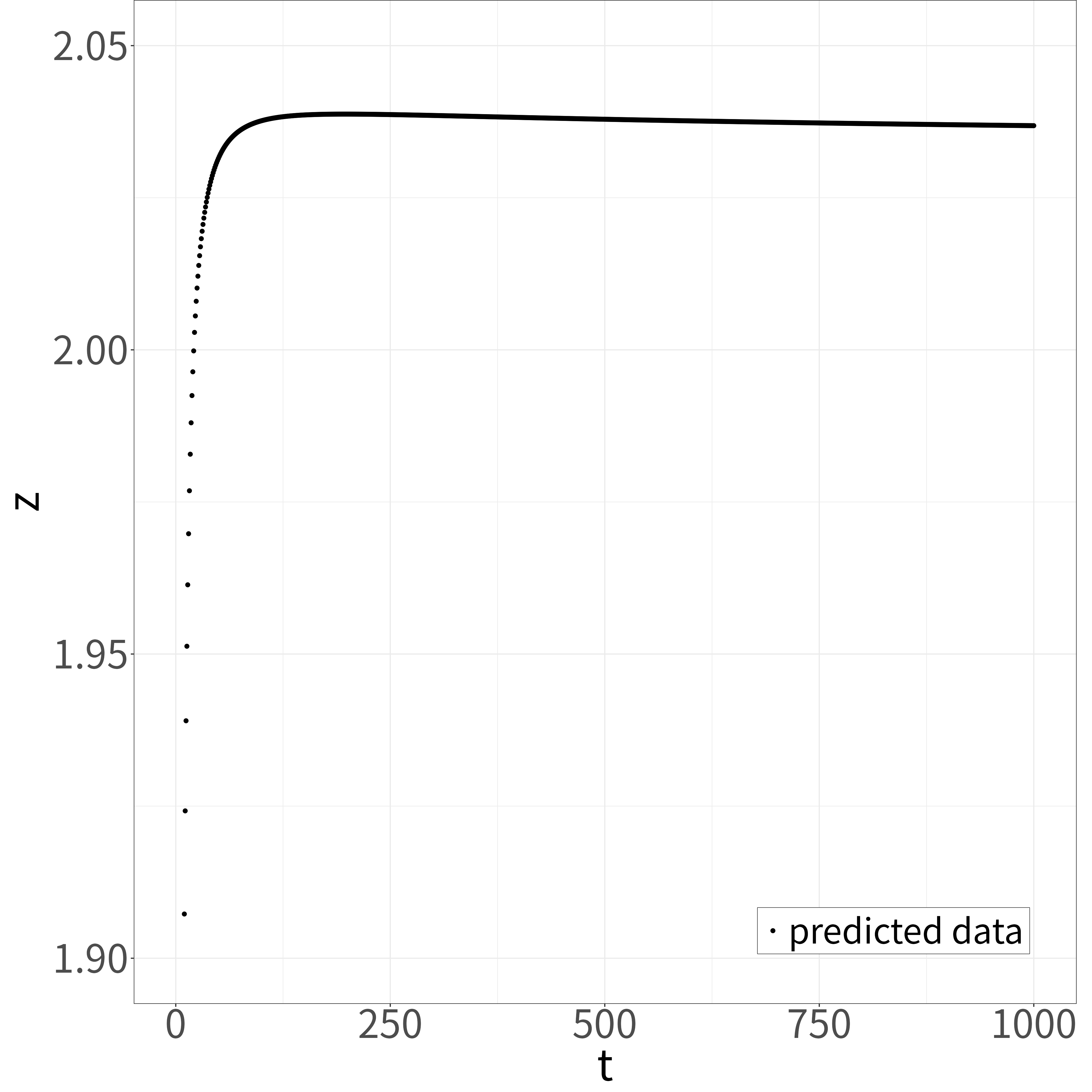}
  \caption{\label{fig:interpolation-z-3D-Ising}
    Interpolation of \(t\) vs \(z\) for the three-dimensional Ising model.
    Black closed circles represent interpolated values at equal intervals of \(t\).
  }
\end{figure}

\Cref{fig:interpolation-beta-3D-Ising,fig:interpolation-gamma-3D-Ising,fig:interpolation-nu-3D-Ising,fig:interpolation-z-3D-Ising} display the interpolations of the local exponents, each plot has \(N = 1003\) data points at equal intervals of \(t\).

Next, we perform the bootstrap method.
We created \(100\) bootstrap samples by resampling \(\Ndata = 100\) data points, which were extracted from the simulation data at equal intervals of \(\log(t)\), \(100\) times.
We applied the present method to each bootstrap sample independently and estimated the mean and numerical error of the critical exponents across bootstrap samples.
Consequently, we estimated the critical exponents to be \(\beta = 0.3252(2)\), \(\gamma = 1.2376(5)\), \(\nu = 0.6293(3)\), and \(z = 2.0346(3)\).
The results are shown in \cref{tab:compare-3D-Ising}.
Compared with the results of the previous NER analysis,~\cite{ito2000nonequilibrium} our results show improved accuracy because of using a stable and automatic method with less influence of human bias.
In the previous NER analysis, a large error bar was necessary because the reliability of convergence decreases due to discrete derivatives.
However, it is considered that the previous analysis provided theoretically correct evaluations.
The present method offers improvements that increase accuracy and reliability.
Our results are close to those in other studies and would be improved if we used more accurate transition temperatures, a larger maximum observation time, or more samples.
Nonetheless, because the present method estimates critical exponents close to the values found in previous studies, we consider it validated.

\begin{widetext}
  \begin{table*}
    \caption{\label{tab:compare-3D-Ising}
    Critical exponents for the three-dimensional Ising model, as obtained from various studies.
    }
    \begin{tabular}{lclllll}
      \hline\hline
      Reference & Method & \(\beta\) & \(\gamma\) & \(\nu\) & \(z\)\\
      \hline
      Ito \emph{et al.} (2000)~\cite{ito2000nonequilibrium,ozeki2007nonequilibrium} & NER(at \(T = 1 / 0.221660\)) & \(0.325(5)\) &  & \(0.635(5)\) & \(2.055(10)\)\\
      Butera and Comi (2002)~\cite{butera2002critical} & High-temperature series &  & \(1.2371(1)\)    & \(0.6299(2)\)   & \\
      Kos \emph{et al.} (2016)~\cite{kos2016precision} & Conformal bootstrap & & & \(0.629971(4)\) & \\
      Ron \emph{et al.} (2017)~\cite{ron2017surprising} & MC renormalization group & \(0.3250(2)\) & \(1.2356(8)\) & \(0.6285(4)\) & \\
      Ferrenberg \emph{et al.} (2018)~\cite{ferrenberg2018pushing} & MC & & \(1.23708(33)\) & \(0.629912(86)\) & \\
      Adzhemyan \emph{et al.} (2022)~\cite{adzhemyan2022dynamic} & \(\varepsilon\) expansion & & & & \(2.0235(8)\) &\\
      Our results  & NER(at \(T = 1 / 0.2216547\)) & \(0.3252(2)\) & \(1.2376(5)\) & \(0.6293(3)\) & \(2.0346(3)\)\\
      \hline\hline
    \end{tabular}
  \end{table*}
\end{widetext}

\section{\NoCaseChange{Summary and Discussion}}
\label{sec:summary-discussion}
We have improved the analysis of fluctuations in the NER method and have applied the present method to estimate critical exponents in both the two-dimensional square Ising model and the three-dimensional cubic Ising model.
The modifications include two significant advancements.
First, we introduced the Gaussian process regression and transformed the physical quantities using \cref{equ:converted-data}.
This transformation enables us to use the data point at \((X, Y) = (0, 0)\) as \(t \to \infty\).
As a result, we can reliably estimate the local exponents \(\beta(t), \gamma(t), \nu(t)\), and \(z(t)\) from simulation data.
Second, under the assumption that these exponents converge exponentially, we extrapolated them using the \(\varepsilon\)-algorithm, which enabled systematic and reproducible extrapolation.
The present method's automation and reproducibility reduce human bias, making it easier to apply the bootstrap method and provide statistical error estimates.
Because the proposed method requires only the data of the relaxation of specific quantities from the Monte Carlo simulation, its simplicity imparts it with strong potential.

The results of our analysis are promising.
For the two-dimensional Ising model, we obtained the critical exponents \(\beta = 0.12504(6)\), \(\gamma = 1.7505(10)\), and \(\nu = 1.0003(6)\), which are consistent with the exact values.
We obtained reliable \(z = 2.1669(9)\) because of the high accuracy of these exponents; this value is consistent with that reported by Nightingale and Bl\"{o}te.~\cite{nightingale2000monte}
The present method offers improvements in accuracy and reliability compared with the previous NER analysis that used discrete numerical derivatives.
These results suggest that the present method is effective at the exact critical temperature.
For the three-dimensional Ising model, we obtained the critical exponents \(\beta = 0.3252(2)\), \(\gamma = 1.2376(5)\), \(\nu = 0.6293(3)\), and \(z = 2.0346(3)\), which are close to the values reported in previous studies~\cite{butera2002critical,kos2016precision,ron2017surprising,ferrenberg2018pushing,adzhemyan2022dynamic} and improved accuracy compared with that achieved in the previous NER analysis.~\cite{ito2000nonequilibrium,ozeki2007nonequilibrium}
These results demonstrate the versatility of the present method and its potential applicability to various models.
Although the results of the present method are not entirely consistent with those of previous studies, its significant advantage lies in its applicability to systems with slow relaxations, such as fully frustrated systems and those undergoing Koserlitz--Thouless transitions.
Therefore, the present method holds promise for analyzing difficult systems and contributing to research on the universality of critical phenomena.

\section*{\NoCaseChange{Acknowledgments}}
\label{sec:orgb6e945a}
The authors are grateful to Kazuaki Murayama for his valuable support and comments.
The authors are grateful to the Supercomputer Center at the Institute for Solid State Physics,
University of Tokyo, for the use of their facilities.
\bibliographystyle{cmpj}
\bibliography{report_fluctuation_jp}
\end{document}